\documentstyle{article}
\everymath{\rm }
\everydisplay{\rm }

\begin{document}
\begin{center}
{\LARGE Energy and directional signatures \\ [.125in]
for plane quantized gravity waves} \\ [.25in]
\large Donald E. Neville \footnote{\large Electronic address:
nev@vm.temple.edu }\\Department of Physics \\Temple University
\\Philadelphia 19122, Pa. \\ [.25in]
March 31, 1997 \\ [.25in]
\end{center}
\newcommand{\E}[2]{\mbox{$\tilde{{\rm E}} ^{#1}_{#2}$}}
\newcommand{\A}[2]{\mbox{${\rm A}^{#1}_{#2}$}}
\newcommand{\Np}{\mbox{${\rm N}'$}}
\newcommand{\Etwo}{\mbox{$^{(2)}\!\tilde{\rm E}\ $}}
\newcommand{\Etld }{\mbox{$\tilde{\rm E}\ $}}
\def \ut#1{\rlap{\lower1ex\hbox{$\sim$}}#1{}}
\newcommand{\phst}{\mbox{$\phi\!*$}}
\newcommand{\psist}{\mbox{$\psi\!*$}}
\newcommand{\bea}{\begin{eqnarray}}
\newcommand{\eea}{\end{eqnarray}}
\newcommand{\be}{\begin{equation}}
\newcommand{\ee}{\end{equation}}
\newcommand{\nn}{\nonumber \\}
\newcommand{\rta}{\mbox{$\rightarrow$}}
\newcommand{\rla}{\mbox{$\leftrightarrow$}}
\newcommand{\eq}[1]{eq.~(\ref{eq:#1})}
\newcommand{\Eq}[1]{Eq.~(\ref{eq:#1})}
\newcommand{\eqs}[2]{eqs.~(\ref{eq:#1}) and (\ref{eq:#2})}
\large
\begin{center}
{\bf Abstract}\\
\end{center}
Solutions are constructed to the quantum constraints for planar
gravity (fields dependent on z and t only) in the Ashtekar
complex connection formalism.  A number of operators are
constructed and applied to
the solutions.  These include the familiar ADM energy and area
operators, as well as new operators sensitive to
directionality (z+ct vs.\ z-ct dependence).
The directionality operators are quantum analogs of the classical
constraints proposed for unidirectional plane waves by Bondi,
Pirani, and Robinson (BPR).  It is argued that the quantum BPR
constraints will predict unidirectionality reliably only for
solutions which are semiclassical in a certain sense.  The ADM
energy and area operators are likely to have imaginary
eigenvalues, unless one either shifts to a real connection, or
allows the connection to occur other than in a holonomy.  In
classical theory, the area can evolve to zero.  A quantum
mechanical mechanism is proposed which would prevent this collapse.
\\[.125in]
 PACS categories: 04.60, 04.30
\section{Introduction: Classical Radiation Criteria}

     The connection-triad variables introduced by Ashtekar
\cite{Ash87} have
simplified the constraint equations of quantum gravity; further,
these variables suggest that in the future we may be able to
reformulate gravity in terms of non-local holonomies rather than
local field operators \cite{RovSmo,GambTri, symmsta}.  However, the
new variables are
unfamiliar, and it is not always clear what they mean physically
and geometrically.  In particular, it is not clear what operators
or structures correspond to gravity waves.  Although the quantum
constraint equations are much simpler
in the new variables, and solutions to these equations have been
found \cite{RovSmo,knotsol, a2}, it is not clear whether any of
these solutions contain gravitational radiation.

     This is the fifth of a series of  papers which search
for operator signatures for gravitational radiation
by applying the Ashtekar
formalism to the problem of plane gravitational waves.  Paper I
in the series \cite{I} constructed classical constants of the
motion for
the plane wave case, using the more familiar geometrodynamics
rather than Ashtekar connection dynamics.  Papers
II and III switched to connection dynamics and
proposed solutions to the quantum constraints \cite{II,III}.   The
constraints  annihilate the solutions of II except at
boundary points, and annihilate the solutions of III everywhere.
Paper IV constructs an operator
$L_Z$ which measures total intrinsic spin around the z axis
\cite{IV}.
The present paper
proposes operator signatures which are sensitive to the
directionality of  gravitational radiation
(z -ct vs.\ z+ct dependence)  and
applies those operators (as well as the spin, energy, and area
operators) to the solutions
constructed in II and III.

     It is not  easy to detect the presence of  radiation, even
when the problem is formulated
classically, using the more familiar metric variables.   One would
like to define gravitational
radiation using an energy criterion (as: radiation is a means of
transporting energy through empty
space ...).   Gravitational energy is notoriously difficult to
define, however, since there is no first-order-in-derivatives-of
-the -metric quantity which is a tensor.   Accordingly, in the
period 1960-1970 several
authors developed an
algebraic  criterion involving transverse components of a
second-order quantity, the Weyl tensor \cite{Petrov, Debvec,
peeling, Penrose} .  To use the criterion, one
needs to know which directions are "transverse"; hence the
criterion is most useful when the direction of propagation is clear
from the symmetry: e.g. radial propagation (for spherical symmetry)
or z-axis propagation (for planar symmetry, the case studied in the
present paper). The planar metrics considered here \cite{EhlK,
Szmet} admit two null
vectors k and l which have the
right hypersurface orthogonality properties to be the propagation
vectors for right-moving (k)
and left-moving (l) gravitational waves along the z axis, so that
the propagation direction  is especially easy to
identify.  The Weyl criterion is derived and discussed in Appendix
D.

     A second, more group-theoretical criterion was developed by
Bondi, Pirani, and Robinson
(BPR) \cite{BPR}.  It is applicable when
the plane wave is unidirectional, that is, when the wave is either
right-moving (depending only on z-ct) or  left-moving (depending
only on z+ct).  The unidirectional case is
especially intruguing.   It is relatively simple, since no
scattering
occurs \cite{Bon, Aich}.   Nevertheless the full complexity of
gravity is already
present; the unidirectional case
is not simply waves propagating
in an inert background.   In particular, no one has been able to
cast the Hamiltonian  into a
free-field form in terms of variables ($\pi_i, q^j$) which commute
in the canonical $[\pi_i,q^j] = -i \hbar \delta_i^j$ manner
characteristic of non-interacting,
non-gravitational theories.   Also, the
BPR criterion for unidirectional radiation requires {\it three}
amplitudes to vanish, rather than
the two one would naively expect from counting the two
polarizations associated with
unidirectional radiation.   I shall argue that the remaining
vanishing amplitude  represents
a constraint on the background geometry, a constraint which must be
satisfied in order for the
waves to propagate without backscattering off  the background.  The
BPR group is derived and discussed in section II.

     Note that one criterion, that based on the Weyl tensor, is
relegated to an appendix; while
the BPR criterion is discussed in the body of the paper.  I have
done this primarily because the
BPR amplitudes are much simpler than the Weyl amplitudes.  It is
not possible to ingnore the
Weyl amplitudes  completely, however;  they are  central to the
literature of the  60's.   Further,
expressions which  appear complex at one time may appear
simple at a later time.  At
one time the
traditional scalar constraint  was thought to be too complex
because it contained a factor of
1/$\sqrt{g}$.    Then Thiemann proposed a regularization of this
constraint which actually
requires that factor \cite{Thie}.  Similarly, the present
"complexity" of the
Weyl amplitudes may disappear
once a quantum regularization is constructed.

     It is of some interest to reexpresss the two  classical
criteria, Weyl and BPR, in the
Ashtekar language, even if one does not go on to consider the
quantum case.  However,
one would really like to construct from each classical criterion a
corresponding quantum operator.  I construct such operators in
section III and argue that
these operators are reliable only when acting on
wavefunctionals for states which are semiclassical, in a sense to
be defined in section III.

     In section IV, I construct additional solutions to the
quantum constraints.  In section V, I apply the BPR quantum
operators to the wavefunctional solutions obtained in
papers II-III, as well as the new solutions constructed in section
IV.  Also in section V, I apply the ADM energy operator to the
solutions, as well as the area operator and the operator $L_Z$
for total intrinsic spin.

     There are three appendices.  Two of the appendices (A
and C) cover calculational
details and the details of the Weyl criterion not used in the body
of the paper.  Appendix B
considers the ADM energy.
There is a modest surprise here: normally the ADM energy is
considered to be
given by the surface term in the
Hamiltonian; but in the quantum case it is possible for the volume
term to contribute also.

          My notation is typical of papers based upon the
Hamiltonian
approach with concomitant 3 + 1 splitup.  Upper case indices A,
B, $\ldots $,I, J, K, $\ldots$ denote local Lorentz indices
("internal" SU(2) indices) ranging over X, Y, Z only.  Lower case
indices a, b, $\ldots $, i, j, $\ldots $ are also three-
dimensional and denote global coordinates on the three-manifold.
Occasionally the formula will contain a field with a superscript
(4), in which case the local Lorentz indices range over X, Y, Z,
T and the global indices are similarly four-dimensional; or a
(2), in which case the local indices range over X, Y (and global
indices over x, y) only.  The (2) and (4) are also used in
conjunction with determinants; e.\ g., g is the usual 3x3
spatial determinant, while\ $^{(2)}e$ denotes the determinant of
the 2x2 X, Y subblock of the triad matrix $e^A_a$. I use Levi-
Civita symbols of various dimensions: $\epsilon _{TXYZ} =
\epsilon _{XYZ} = \epsilon _{XY} = +1$.  The basic variables of
the Ashtekar approach are an inverse densitized triad \E{a}{A}
and a complex SU(2) connection \A{A}{a}.
\begin {eqnarray}
     \E{a}{A}& =& e e^a_A; \\
\label{eq:1.1}
     [\E{a}{A},\A{B}{b}]&=& \hbar \delta (x-x') \delta ^B_A
\delta ^a_b.
\label{eq:1.2}
\end{eqnarray}

     The planar symmetry (two spacelike commuting Killing
vectors, $\partial_x$ and $\partial_y$ in appropriate coordinates)
allows Husain and Smolin \cite{HSm} to
solve and eliminate four  constraints (the
x and y vector constraint and the X and Y Gauss constraint) and
correspondingly eliminate four pairs of  (\E{a}{A}, \A{A}{a})
components.  The 3x3 \E{a}{A}\
matrix then assumes a block diagonal form, with one 1x1 subblock
occupied by \E{z}{Z}\, plus one 2x2 subblock which contains all
the ``transverse''  \E{a}{A}, that is, those  with a = x,y and A =
X,Y.  The
3x3 matrix of connections
\A{A}{a}\ assumes a similar block diagonal form.  None of the
surviving fields depends on x or y.

     The local Lorentz indices are vector rather than spinor;
strictly speaking the internal
symmetry is O(3) rather than SU(2), gauge-fixed to O(2) rather than
U(1).  Often it is convenient
to shift to transverse fields which are eigenstates of the
surviving gauge invariance O(2):
\begin{equation}
     \E{a}{\pm} = [\E{a}{X} \pm i\E{a}{Y} ]/\sqrt{2} ,
\label{eq:1.3}
\end{equation}
where a = x,y; and similarly for \A{\pm}{a}.

     In papers I-III I use the letter H to denote a constraint
(scalar, vector, or Gauss).  In the
present paper I adopt what is becoming  a more common convention in
the literature and use the
letter C to denote a constraint, while reserving the letter H for
the Hamiltonian.  The quantity denoted $C_S$ in the present paper
is identical to the constraint denoted $H_S$ in papers II-III.
This convention
underscores  the fact that every gravitational theory has
constraints, but not every gravitational
theory has a Hamiltonian.

     In three spatial dimensions it is usual to place the
boundary surface at spatial infinity.  Bringing the surface at
infinity in to finite points is a major change, because at
infinity the metric goes over to flat space, and flat space is a
considerable simplification.  In the present case (effectively
one
dimensional because of the planar symmetry)  the space does {\it
not} become flat at z goes to infinity, and nothing is lost by
considering an arbitrary location for the boundary surface.  The
``surface'' in one dimension is of course just two points (the two
endpoints of a
segment of the z axis).  The notation $z_b$ denotes either the left
or right boundary point $z_l$ or $z_r$,
$z_l \leq z \leq z_r$.  The result that the space
does not become flat as z goes to infinity was established in paper
II.  Note that this result agrees
with one's  intuition from Newtonian gravity, where the potential
in one spatial dimension due to a bounded source does
not fall off, but grows as z at large z.

     If a certain solution does not satisfy the Gauss constraint
(or other constraint) at the boundary, this does not mean that
necessarily there is something wrong with the solution.  In
classical theory the solutions satisfy the constraints everywhere.
In quantum theory, however, when the constraints are imposed after
quantization, in the Dirac manner, it is only necessary that the
{\it smeared} constraint annihilate the solution:
\be
     \int dz \delta N(z) C(z) \psi = 0.
\label{eq:1.4}
\ee
The expression $\delta N C$ generates a small change $\delta N$ in
the Lagrange multiplier N.  If N obeys a boundary condition of the
form $N \rta $ constant at boundaries $z_b$, then \eq{1.4} must
respect this boundary condition, which means
\be
     \delta N(z_b) = 0.
\label{eq:1.5}
\ee
\Eq{1.5} implies that $C(z_b) \psi $ does not have to vanish.
A statement that "this solution does not obey the constraint at
the boundaries" does not mean necessarily that the solution is
flawed.

\section{Bondi-Pirani-Robinson Symmetry}

     Bondi, Pirani, and Robinson (BPR) argue that the metric of a
unidirectional plane gravitational wave should be invariant under
a five-parameter group of symmetries.  Their argument proceeds
essentially as follows.  First they point out that a
plane {\it electromagnetic} wave moving in the +z direction is
invariant under a five parameter group.  (Besides the obvious
$\partial_x$, $\partial_y$, and $\partial_v$ symmetries, there are
two "null
rotations" which rotate the $v= (t+z)/\sqrt{2}$ direction into
the
x or y direction.)   Then for gravitational plane waves they
construct five Killing vectors which have the same Lie algebra as
the corresponding Killing vectors for the electromagnetic case.
(More precisely they construct
ten Killing vectors, one set of five for ct+z waves, and a similar
set of five for ct-z waves.)

     This
section constructs the five ct-z vectors, then imposes the usual
symmetry requirement that the
Lie derivative of the basic Ashtekar fields must vanish in the
direction of the Killing vectors.   In
this way one finds that the fields must obey certain constraints;
the section closes with a
discussion of the physical meaning of these constraints in the
classical theory.

     It is convenient to do the proofs in a gauge which has been
simplified as much as possible
using  the $\partial _v$ symmetry, then afterwards transform the
results to a general gauge.  The
plane wave metrics we consider here possess two hypersurface
orthogonal null vectors; if the
two hypersurfaces are labeled u = constant and v = constant (u and
v = $(ct \pm z)/\sqrt{2}$),
then one can always transform the metric to a conformally flat form
in the (z,t) sector by using u
and v as coordinates \cite{Szmet}:
\be
      ds^2 = -2dudv f(u,v) + \sum ^{(2)}g_{ab} dx^a dx^b.
\label{eq:3.1a}
\ee
The sums over a,b,c,$\ldots$ extend over x,y only.   If one now
invokes the symmetry under
$\partial _v$, then f(u,v) depends on u only, and one can remove
the function f by transforming to a
new u coordinate.   In this gauge,
Rosen gauge \cite{Rosmet}, the metric is (not just conformally
flat, but) flat in the (z,t) sector
and non-trivial only in the (x,y) sector.
\be
     ds^2 = -2dudv + \sum ^{(2)}g_{ab} dx^a dx^b.
\label{eq:3.1}
\ee
In Rosen gauge, the Killing vectors are $\partial_x$,
$\partial_y$, $\partial_v$, and
\be
     \xi^{(c)\lambda} = x^c \delta^{\lambda}_v + \int^u
g^{cd}(u')du'\delta^{\lambda}_d.
\label{eq:3.2}
\ee

     The constraints imposed by the first three $\partial_x$, $
\partial_y$, and $\partial_v$  Killing vectors are satisfied
already, because of the
choice of gauge.   I
now work out the constraints
which the last two Killing vectors,  \eq{3.2}, impose on the
Ashtekar variables (in Rosen
gauge first, than in a general gauge).  I
summarize the highlights of the calculation in this section, and
move the algebraic details to
appendix A.

     It is necessary to calculate the symmetry
constraints on the four-dimensional
tetrads and Ashtekar connections first, since the Killing vectors
are intrinsically four-dimensional, then carry out a 3+1
decomposition to
obtain the constraints on the usual three dimensional densitized
triad and connection.   At the four-dimensional level,  the three
local
Lorentz boosts have been gauge-fixed
by demanding that three of the tetrads vanish:
\be
      e^t_M = 0, M = space.
\label {eq:3.3}
\ee
The gauge condition of \eq{3.3} is the standard choice, used with
all metrics \cite{Pel}.  In addition,
for the special case of the plane wave metric, the gauge-fixing
of the XY Gauss constraint and
xy spatial diffeomorphism constraints imply that four more
tetrads vanish \cite{HSm}.
\be
     e^z_X = e^z_Y = e^x_Z = e^y_Z = 0.
\label{eq:3.4}
\ee

     At the four-dimensional level, the
requirement of vanishing Lie derivative in the
direction of the Killing vector  gives
\bea
     0 &=& \xi^{\lambda} \partial_{\lambda}e^{\alpha}_I -
               \partial_{\beta}\xi^{\alpha} e^{\beta}_I -
               L^{I'}_{.I}  e^{\alpha}_{I'};
\label{eq:3.5a} \\
     0 &=& \xi^{\lambda} \partial_{\lambda}^{(4)}A^{IJ}_{\alpha}
               + \partial_{\alpha}\xi^{\lambda}\,
               ^{(4)}A^{IJ}_{\lambda} \nn
       & & + {\cal L}^I_{.I'}\, ^{(4)}A^{I'J}_{\alpha} +
               {\cal L}^J_{.J'}\, ^{(4)}A^{IJ'}_{\alpha} -
               \partial_{\alpha}{\cal L}^{IJ}.
\label{eq:3.5b}
\eea
These equations are not quite the usual Lie derivatives because
of the the L and ${\cal L}$ terms.
L and $ {\cal L}$ are local Lorentz transformations.  If $\xi =
\partial_x,\  \partial_y,\ or\ \partial_v$, no
L or ${\cal L}$ is required.  If  $\xi = \xi^{(c)}$,
one of the two Killing vectors defined at \eq{3.2}, then a
Lorentz transformation L is
required in \eq{3.5a}; otherwise  the symmetry  destroys
the gauge conditions, \eqs{3.3}{3.4}.    Then for
consistency,
$^{(4)}A$ in \eq{3.5b} must undergo the same Lorentz
transformation;
since $^{(4)}A$ is self-dual,
the Lorentz transformation ${\cal L}$ in \eq{3.5b} must be the
self-dual version of the Lorentz
transformation L:
\be
     2{\cal L}^{IJ} = L^{IJ} + i\delta (\epsilon^{IJ}_{..MN}/2
                         \epsilon_{TXYZ})L^{MN}.
\label{eq:3.6}
\ee
The phase $\delta /\epsilon_{TXYZ}= \pm 1$ is the duality
eigenvalue which determines
whether the theory is self-dual or anti-self-dual.  Because I
include the extra factor of
$\epsilon_{TXYZ}$,  \eq{3.6} contains two factors of $\epsilon$, so
is independent of one's
choice of phase for this quantity.  After the four-dimensional
theory is rewritten in 3+1 form, all
results  will depend only on $\delta$.  In the body of the paper I
choose $\delta = +1$, but
appendix A indicates what happens for the opposite choice $\delta
= -1$.

     It is a straightforward matter to determine the Lorentz
transformation L which will preserve the gauge conditions of
\eqs{3.3}{3.4}, then solve \eqs{3.5a}{3.5b}.  This is done in
appendix A.  From \eq{a.20} in that appendix, \eqs{3.5a}{3.5b}
imply the
following constraints on the connection A.
\bea
     0 & =& A^{-}_a ; \nn
     0 & = & - A^{+}_a + 2 \mbox{``Re''} A^{+}_a, \nn
     & & \mbox{(right-moving)}
\label{eq:3.7a}
\eea
where a = x,y only.  The connection A is now the usual 3+1
connection, not the four-dimensional connection $ ^{(4)}A $.  Re
$ A^{X}_a$ without the quotes
is the usual real part, containing no factors of
i, while
\be
     \mbox{``Re''} A^{+}_a \equiv (\mbox{Re} A^{X}_a
                        +  i \mbox{Re} A^{Y}_a)/\sqrt{2} .
\label{eq:3.7b}
\ee
``Re'' A contains a factor of i, because of the i in the definition
of the ($X \pm i Y$)  O(2) eigenstates,
and is no longer real.  If one writes out the ``Re'' and ``Im''
parts of
$ A^{\pm}_a $ it is easy to see that the two constraints of
\eq{3.7a} are just the complex conjugates of each other.  To obtain
the constraints for
left-moving waves, interchange + and - in \eq{3.7a}.

     \Eq{3.7a} can be interpreted physically by using the classical
equations of motion to prove  theorems about the
spin behavior of the BPR fields.  Again, the required
calculations  are done in an appendix (appendix B); and this
section  summarizes the main conclusions.

     To interpret the spin content of the four
amplitudes which vanish, it is better to work with the four
combinations \E{a}{A} \A{-}{a} and \E{a}{B} [-\A{+}{a} + 2 ``Re''
\A{+}{a}].   (One can always recover the original four amplitudes
from these four, because one can always invert the 2x2 matrix
formed from the transverse
components of the
densitized triad, \E{B}{b} with B = X,Y and b = x,y.)  \Eq{c16} of
appendix B  expresses
the total spin angular momentum $ L_Z $ of the gravitational
wave in terms of these combinations.
\bea
   L_Z& =& i \int dz \{ e^y_+ e_{+x} \E{a}{-}[\A{-}{a} + (\A{-}{a}
- 2 Re \A{-}{a}) \nn
          & & + e^y_- e_{-x} \E{a}{+} [\A{+}{a} + (\A{+}{a}
          - 2 Re \A{+}{a}) \} - (x  \rla y),
\label{eq:3.8}
\eea
where  $ e_{Aa} $ and  $ e^a_A $ are triad and inverse triad
fields, respectively.  Out of the four possible combinations
\E{a}{A} \A{-}{a} and \E{a}{B}
[-\A{+}{a} + 2 ``Re''
\A{+}{a}], only  two combinations \E{a}{-} \A{-}{a} and
\E{a}{+} [-\A{+}{a} + 2
``Re'' \A{+}{a}] contribute to the spin angular momentum.  Since
these are the two
amplitudes with O(2) helicity $ \pm $ 2 in the local Lorentz
frame, it is natural to interpret $ \E{a}{\pm} \A{\pm}{a}$ as an
amplitude for a wave having helicity $ \pm $2.  Both  helicity $
\pm $2 combinations must
vanish in order to eliminate the two  polarizations moving in the
ct + z direction.

      Two more, helicity zero
combinations, $  \E{a}{+} \A{-}{a} $ and $  \E{a}{-} [-\A{+}{a} +
2
\mbox{``Re''} \A{+}{a}] $, also contain the fields of
\eq{3.7a}.   How does one interpret these helicity zero amplitudes?

Using the Gauss constraint plus the
classical equations of motion in a conformally flat gauge,
one can prove that these two constraints collapse to a single
constraint, \eq{c20} of Appendix B.
\be
     0 =  (\partial_t + \partial_z ) \E{z}{Z} .
\label{eq:3.9}
\ee
\E{z}{Z} = $e^z_Z e$ = $ ^{(2)}e $, where $ ^{(2)}e $ is the
determinant of the 2x2 transverse
sector of the triad matrix, a scalar function of (z,t).   This
function would seem to characterize
the background geometry, rather than the wave.  In general
relativity,  however, ``background''
and wave are inseparable, in the sense that the ``background''  is
not inert.   The wave will
scatter off the background, in general, unless it obeys the
constraint given by \eq{3.9}.

     This completes the survey of the constraints predicted by BPR
symmetry, and their physical interpretation in the classical
context.  In the next section these classical expressions are
promoted to quantum operators.

\section{The Transition from Classical to Quantum Criterion}

     This section lists four issues which arise when converting a
classical expression into a
quantum criterion.  I summarize and discuss each issue, then show
the application to the  BPR
criteria.

\subsection{Factor ordering and regularization}

     I choose a factor ordering which is natural and simple
within the complex
connection formalism.  The ordering is ``functional derivatives
to the right''\cite{OptrOrd}.  That is,  I quantize in a standard
manner, by replacing one
half the fields by functional derivatives,
\bea
     \E{z}{Z} &\rta& -\hbar \delta / \delta \A{Z}{z}; \nonumber
\\
     \A{A}{a} &\rta& +\hbar \delta / \delta \E{a}{A}, \nonumber
\\
               & &  \mbox{(for a = x,y and A = X,Y)},
\label{eq:2.0}
\eea
and then order the functional derivatives to the right in every
operator or constraint.
Regularization, if needed, is via point splitting \cite{HSm}.  This
approach
has the virtue of consistency,
since I have used it in two previous papers on quantization of
plane waves.

\subsection{Semiclassicality}

       An example from QED will be helpful in
explaining what is meant by
semiclassicality.  The
BPR criteria are essentially field strengths for waves moving in
a
given direction with  a given
polarization.  An analogous quantity from  flat space QED  is
\be
     {\cal F} \equiv {\cal F}^{\mu \nu} m_{\mu} k_{\nu}
\label{eq:2.1}
\ee
Here ${\cal F}^{\mu \nu} $ is the self-dual QED field strength
and
(k,l,m,\={m }) is the usual
flat space null tetrad: k and l are null vectors with space
components along $\pm $ z ; m and \={m} are
transverse polarization vectors along $(\hat{x} \pm
i\hat{y})/\sqrt{2}$.    Classically, the criterion
for absence of  radiation along k with polarization \= {m} is
${\cal F}$ = 0.  The
corresponding
quantum criterion for absence of radiation, obtained after
replacing classical A fields by
quantum operators  \^{A} is
{\it not}
\be
     \hat {\cal F} \psi = 0,     \mbox {(wrong)}
\label{eq:2.2}
\ee
but rather
\be
     \langle semicl|\hat{\cal F}|semicl \rangle = 0.
\label{eq:2.3}
\ee
Since the QED field strength $\hat{\cal F}$ contains creation as
well
as annihilation operators, it
cannot
annihilate any state, and \eq{2.2} is too strong.  The classical
statement ${\cal F}$ = 0
merely implies the existence of a corresponding semiclassical
state
such that  \eq{2.3} holds.  I have deliberately used the
term ``semiclassical''  rather than
``coherent'  to describe the state in \eq{2.3}, because the
latter
term conventionally denotes  a
state which is an eigenfunction of the annihilation operator, and
annihilation operators usually
are not available in quantum gravity.  While annihilation
operators
may not exist, certainly
semiclassical states will, because of the correspondence
principle.

     Clearly the criterion \eq{2.3} is more difficult to apply
than
\eq{2.2}, but let us survey the
damage;  the situation may not be hopeless.  To define
``semiclassical''   without invoking
coherent states or annihlation operators, one can study
\be
     \hat{F}|semicl \rangle = f(z)|semicl \rangle
                                   + |remainder \rangle,
\label{eq:2.4}
\ee
where \^{F} stands for a typical BPR field.  $|semicl \rangle$ is
normed to unity, although $|remainder \rangle$ need not be.
I define semiclassical, not by requiring f(z)  to be large, but
rather by requiring the
$|remainder \rangle $ to be small.   If I require f(z) to be
large
(perhaps
reasoning that ``classical''  means
large quantum numbers) then I exclude the vanishing amplitude
case,
f(z) = 0, where $|semicl \rangle $ is
the vacuum with respect to a given radiation mode.  The vacuum is
a well-defined state,
classically, and one expects it to have a quantum analog.  For
$|semicl \rangle$  to approximate a
classical state, therefore, it is not necessary that f(z) be
large;
only that the fluctuations away
from this state be small.  These fluctuations are measured by
\be
     \langle \hat{F}\dagger \hat{F} \rangle - \langle
\hat{F}\dagger \rangle \langle \hat{F} \rangle =
          \langle remainder|remainder \rangle  \leq
               (l_p/l)^2/l^2.
\label{eq:2.5}
\ee
Here the fluctuations are assumed to be small
compared to the size of typical matrix elements $\langle
f|\hat{F}|i \rangle
\approx l_p/l^2$ one gets when $|i\rangle$ and $\langle f|$ are
few-graviton states and \^{F} is a canonical degree of freedom in
the linearized theory.   $l_p$ is the
Planck length, and l is  a typical length
or wavelength.  (To modify
this discussion  for the Weyl fields of appendix D, which are
classical dimension
$1/l^2$, replace $l_p/l^2$ by
$l_p/l^3$.   From now on $\langle \ \rangle$ is understood to
indicate an average
over a
semiclassical state, unless
explicitly indicated otherwise.)

     One might ask what is meant by a typical length l, in a
quantized and diffeomorphism
invariant theory where no background metric is available.  In
such
a theory, even in the absence
of a background metric, length, area, and
volume operators can be defined \cite{AV, Lollvol, Thie}, and  the
eigenvalues of
these
geometric operators are
dimensionless functions of spins $j_i$, times  factors of $l_p$
to give the correct
dimension.  The spins label the irreducible representations of
SU(2) associated with each
holonomy (if the wavefunctional is in connection representation,
a product of holonomies) or
associated with each edge of a spin network (if the
wavefunctional is a spin network state).
Thus one expects $l = l_0(j_i) l_p$, $l_0$ dimensionless and $\gg
1$.  (In the planar case, SU(2) is gauge-fixed to O(2) and
presumably the SU(2) eigenvalues j will be replaced by the O(2)
eigenvalues m = spin angular momentum component along z.)
Evidently, then, to check
that \eq{2.5} is satisfied, one should apply the geometric
operators to the state first, in order to
estimate  l.  One also needs the
measure in Hilbert space; but
in favorable situations one might be able
to tell that $|remainder \rangle$ is small simply by inspection
of
\eq{2.4}.

     Since the criterion of \eq{2.5} is an inequality, it cannot
be
used to draw sharp distinctions
between states.  For example, in the linearized limit,
if $|N \rangle$
denotes an eigenstate of the number
operator having N quanta of a given polarization and direction,
then f(z) will be zero, while the
norm \eq{2.5} will be order N $(l_p/l)^2/l^2$.  There will be
uncertainties in estimating l, so that
the criterion cannot distinguish sharply between the  vacuum
state
and a number eigenstate
having small occupation number N.

\subsection{Non-polynomiality}

      The  BPR operators, \eq{3.7a}, occur in complex
conjugate pairs, and one
member of the BPR pair involves Re \A{A}{a}, which is a known,
but
non-polynomial function
of  \E{a}{A}.  In
particular, Re \A{A}{a} contains  factors of
1/\Etwo, where \Etwo\  is the 2x2  determinant formed from the
\E{a}{A} with  internal indices A
= X,Y and global indices a = x,y.  I have dealt with a similar
operator, 1/\E{z}{Z}, in a previous
paper; but would just as soon not do so here.

      One can use the fact that the BPR constraints come in
complex cojugate pairs, plus
semiclassicality, to prove the following theorem (and then one
uses the
theorem to avoid dealing with
the non-polynomiality). Theorem:
\be
      \langle - \A{+}{a} + 2 Re \A{+}{a} \rangle
           = \langle \A{-}{a} \rangle *,
\label{eq:2.0b}
\ee
and similarly for the other BPR pair.  This result is just what
one
would expect from the
corresponding result for expectation values of complex operators
in
ordinary quantum
mechanics (for instance $\langle p+iq \rangle * = \langle p-iq
\rangle$) except that here the
basic operators are not
Hermitean, so
that the proof is slightly longer.  Proof: Expand out the
$\A{\pm}{a} $ operators using
\bea
     \A{\pm}{a}& =& (\A{X}{a} \pm i\A{Y}{a})/\sqrt{2} \nonumber \\
         &=& \hbar(\delta/\delta \E{a}{X} \pm i\delta/\delta
                         \E{a}{Y})/\sqrt{2}.
\eea
Integrate by parts the functional derivative on the left side of
\eq{2.0b}, using
\be
     \int \mu \psi* \hbar \delta \psi /\delta \E{a}{A}=
           \int \mu [(- \hbar) \delta/\delta \E{a}{A} \psi *]
\psi
+
           \int \mu \psi * 2 Re \A{A}{a} \psi.
\label{eq:2.0a}
\ee
Here $\psi$ is the semiclassical state and $\int \mu $ is the
measure, a path integral over the
fields in $\psi$. $\mu$ need not be known in detail, except that
it
enforces the reality condition
in \eq{2.0a} ( via $(- \hbar) \delta \mu /\delta \E{a}{A}=  2 Re
\A{A}{a} \mu$).  Also, $\mu $
must be real ( $\mu * = \mu$)  in order for norms to be real .
If
one inserts the relations
\eq{2.0a} on
the left in \eq{2.0b} and carries out the complex conjugation
(using $\mu * = \mu$); the result
is
the right-hand side of \eq{2.0b}.$\Box$

\subsection{Kinematic  vs. physical operators}

     I continue with the
comparison of  the BPR operators
to field strengths in QED.  In QED, operators representing
observables should  involve only true,
rather than gauge degrees of freedom, therefore should commute
weakly with the smeared
Gauss constraint.   Similarly, in quantum gravity, observables
should commute with the
constraints.  The BPR operators do not do this.

     An operator will be said to be
kinematic if it commutes only with the Gauss and   diffeomorphism
constraints; and physical if
it commutes with all the constraints, including the scalar
constraint.  It is desirable
for  an operator to commute with as many of the Hamiltonian
constraints as possible, since the
operator is then presumably independent of one's choice of
arbitrary coordinates.  Physical
operators are best of all, but even a diffeomorphism and Gauss
invariant operator can furnish
valuable information about the physical meaning of a state.  The
volume, length, and area
operators, for example,  are only diffeomorphism and Gauss
invariant.

     I shall show below that the BPR amplitudes for the
unidirectional case can be modified so as
to make them
kinematical.  I shall refer to the modified amplitudes as the
unidirectional BPR amplitudes, since
they continue to refer to excitations
moving in a single direction.  (That is, it is not necessary to
combine a right-moving operator
with a left-moving operator to get a kinematical operator.)
Further, in both the scattering and
unidirectional cases, it is possible to construct additional
combinations of the BPR operators
which are actually physical, rather than kinematic only;
but (except in the linearized limit) these physical  combinations
no longer refer to excitations
moving in a single direction.

     From the remarks at the end of section I, in the planar
symmetry case only three constraints
$\hat{C}_i$ survive gauge fixing: the generator of spatial
diffeomorphisms along
z, the Gauss constraint for internal rotations around Z, and  the
scalar constraint.  Consider first $\hat{C}_z$, the
diffeomorphism
constraint.  The BPR fields are
local, scalar functions of (z, t), of density weight zero, and
therefore
do not commute with $\hat{C}_z$.  Evidently the BPR fields have
unacceptable dependence on
the arbitrary coordinate label z.  The simplest remedy is to get
rid of z by
integrating the BPR amplitude $\hat{F}$, over z, following the
prescription that observables in quantum gravity should be
non-local; but the density weight of $\hat{F}$
must be unity rather than zero in order for integration to
produce
a diffeomorphism invariant quantity.
One would like a density weight of unity for another reason:
operators which are not density
weight unity often  turn out to be background-metric-dependent
when
regulated \cite{diffreg}.

     There are two ways to  modify the BPR operators to make them
density weight unity.
(Either) replace
\bea
     \A{\pm}{a} &\rta& \E{a}{\pm'}\A{\pm}{a};
                    \nonumber \\
     -\A{\mp}{a} + 2 Re \A{\mp}{a}&\rta& \E{a}{\pm'}
               [-\A{\mp}{a} +  2 Re \A{\mp}{a}];
                    \mbox{a = x,y only};   \nonumber \\
          & & \mbox{(unidirectional)}
\label{eq:2.9a}
\eea
(or) replace
\bea
 \A{B}{b} &\rta& \E{a}{B}\A{B}{b};
               \nonumber \\
     -\A{B}{b} + 2 Re \A{B}{b}&\rta& \E{b}{B}
               [-\A{B}{a} +  2 Re \A{B}{a}]; \mbox{a,b = x,y
only};
               \nonumber \\
          & & \mbox{(physical)}
\label{eq:2.9b}
\eea
Classically, both these transformations are invertible.  (Recall
that the \E{a}{A} matrix is block
diagonal,
with one 2x2 subblock containing a = x,y, and A = X,Y or +,-
only.)
Both sets of  new operators
contain the same
information as the old, therefore; and I can adopt either set as
the new quantum BRS
criterion  (after integration over dz and inclusion of appropriate
holonomies to secure gauge
invariance; see below).  The first set is easier to interpret,
since each
operator contains only \A{+}{a} or only
\A{-}{a} but not both, hence the name ``unidirectional'': each
operator refers to either left-
or right-moving excitations, as
do the original BPR operators.  Each operator in the second set
is
a mix of left- and right-moving
operators.  The first set, though not  gauge
invariant, can be made kinematic by sandwiching between appropriate
holonomies and
intgrating over z.  The second set needs
only an integration over z to make it (not only kinematical, but
also) physical.  I now verify the foregoing two statements in
detail.

      The first set can be made Gauss invariant by sandwiching
each
operator between
holonomies.    Since the internal rotations
around X and Y are fixed, and only rotations  around Z remain,
the
gauge
group is U(1) rather than
SU(2).  The irreducible representations are one-dimensional and
labeled by a single integer or
half-integer, the spin along Z.  In \eq{2.9a} some of the new
unidirectional BPR
fields have spin $\pm + \pm' = 0$, and are already Gauss
invariant, while others have
spin
$\pm + \pm' =\pm2$; only  the latter
fields
are non-invariant and need to be
sandwiched between holonomies, for example
\bea.
      \E{a}{+}\A{+}{a}& \rta& \int dz M(z_r,z) \E{a}{+}(z) S_-
                                        \A{+}{a}S_-
                         \nonumber \\
                    & & \times  M(z,z_l) M(z_l, z_r)
                         \nonumber \\
                    & &  \mbox{(unidirectional) },
\label{eq:2.10}
\eea
where M is a holonomy along the z axis,
\be
     M(z_2,z_1) = exp[i\int_{z1}^{z2} S_Z \A{Z}{z}(z') dz'.
\label{eq:2.11}
\ee
 $S_{\pm}, S_z$ are the usual Hermitean SU(2) generators.  In
\eq{2.10}the
integration over z has been added to enforce spatial
diffeomorphism
invariance.  If
$M(z_2,z_1)$ is pictured as a
flux line running from $z_1$ to $z_2$. then  \eq{2.10} contains
two
parallel flux lines forming
a hairpin contour, with
the open end at the right-hand boundary $z_{r}$: reading from
left to right, each flux line starts
at $z_{r}$, runs to
\E{a}{+} \A{+}{a}, then to $z_{l}$, the closed end of the
hairpin; then the flux line
turns around and returns to $z_{r}$.  The flux line must be
pictured as ``open''  at the $z_{r}$ end, because the
$S_{-}S_{-}$
in the middle of the contour
changes the eigenvalue of the $S_z$ in the holonomies.  For the
ingoing holonomy
$M(z_{r},z)$, $S_z$ in
\eq{2.11} is evaluated at one eigenvalue, $S_z = m_z$; while for
the return holonomy
$M(z_{r}, z_{l})$, $S_z$ is evaluated at a different eigenvalue
$S_z = m_z +2$.  At the open
end of the flux line, the $m_z$ values do not match, and there is
no possibility of taking a trace.

     The operators of \eq{2.10} are gauge invariant,
despite the absence of a trace  and the
presence of open  flux lines at z =
$z_{r}$.  (This is the point in the argument where one uses the
assumption of
unidirectionality.)  The wave   is travelling toward $z_{r}$ but
has not yet
reached that point; therefore
the boundary condition at $z_{r}$ is flat space.  Internal Gauss
rotations must respect this
boundary condition, which means the triad at $z_{r}$ cannot be
rotated.  Technically, the
Gauss constraint $C_G$ must be smeared, $\int \Lambda (z) C_G(z)
dz$; and the smeared
constraint commutes with the operators in \eq{2.10} at $z =
z_{r}$, because $Lim_{z \rta
z_{r}} \Lambda (z) = 0$.

     It is
straightforward to rewrite the holonomies as exponentials to the
$m_z$ or $m_z +2$ power,
and convince onself that the value of the constant $m_z$ does not
matter.  Also  one can use any
(2j+1)x(2j+1) dimensional representation of the matrices $S_{-}$,
$S_z$, and the value of j
does not matter: changing j merely changes the operators in
\eq{2.10} by constants.

     One can also check that the  criterion of
\eq{2.9b}
is physical, after  integration
over dz.
\be
     G^a_b \equiv \int dz \E{a}{B} \A{B}{b}.
\label{eq:2.10b}
\ee
(Husain and Smolin \cite{HSm} use the notation $K^a_b$ for these
operators,
but it is perhaps better to reserve the letter K for extrinsic
curvature.)  For the most part the check  is a straightforward
computation of the commutator
between the criterion and the scalar constraint; and the proof has
been done already by Husain and Smolin.  I will discuss
only the one step which involves
a slight subtlety, an integration by parts.  (This step is trivial
in the case studied by Husain and Smolin, since their z axis has
the topology of a circle and there are no surface terms.) One
starts from  the
integrated scalar constraint  $\int dz' \delta \ut{N} (z')
\hat{C}_S(z')$ , where
\be
    \hat{C}_S = \E{c}{C}\E{d}{D} \epsilon_{cd}
         \epsilon^{CD}\A{M}{m}\A{N}{n}\epsilon^{mn}\epsilon_{MN}/4
          +\epsilon_{CD}\E{c}{C}F^{D}_{zc} \E{z}{Z}.
\label{eq:2.13}
\ee
$\epsilon_{cd}$ is the two dimensional constant Levi Civita
symbol
(c,d = x,y or X,Y depending
on context), and F is the field strength
\be
     F^D_{zc} = (\partial_z \delta_E^D - \epsilon_{DE}
                    \A{Z}{z})\A{E}{c}.
\label{eq:2.14}
\ee
$\delta \ut{N} $ is a small change in the densitized lapse.  Since
it  must leave invariant  the
coordinate system at the boundaries, it  vanishes at boundaries.
In II-III I used a scalar constraint which equals \eq{2.13}
divided
by a factor of \E{z}{Z}, but
here I use  \eq{2.13} itself, for simplicity, because factors of
\E{z}{Z} do not matter: the
operators of \eq{2.9b} commute with \E{z}{Z}.
The canonical commutator is
\be
     [\A{A}{a}, \E{b}{B}] = \hbar \delta ^A_B \delta ^b_a \delta
(z-z'),
\label{eq:2.15}
\ee
and one must use this to evaluate
\be
     [\int dz \E{a}{B}\A{B}{b}, \int dz' \delta \ut{N}(z')
\hat{C}_S(z') ].
\label{eq:2.16}
\ee

$\hat{C}_S$, \eq{2.13}, is the sum of two terms, and only the
commutator with the second
term involves an integration by parts.  The second term is
proportional to  the field strength $
F^D_{zc}$, so that the
commutator has the form
\bea
      \int dz dz' \delta \ut{N}(z') &\cdots&
[\E{a}{B}\A{B}{b}(z),
                    \E{c}{C}F^{D}_{zc}(z') ]
                              \nonumber \\
      = \int dz dz' \delta \ut{N}(z')& \cdots
&\{\E{c}{C}[\E{a}{B},
F^{D}_{zc}]\A{B}{b}
                    +\E{a}{B}[\A{B}{b}, \E{c}{C}]F^{D}_{zc}\}
                              \nonumber \\
     = \int dz dz' \delta \ut{N}(z')& \cdots& \hbar
\{\E{a}{C}(z')[-\partial_{z'} \delta (z-z')
               \A{D}{b}(z) +\epsilon_{DB}\A{Z}{z}\A{B}{b}] \nn
             & &  +\E{a}{C}[\delta (z-z')]F^{D}_{zb}\}
                                         \nonumber \\
    = \int dz dz' \delta \ut{N}(z')& \cdots& \hbar
          \{\E{a}{C}(z')[-F^{D}_{zb}\delta (z-z')] \nn
     & &   +\E{a}{C}[\delta (z-z')]F^{D}_{zb}\} + ST
                          \nonumber \\
    = 0 + ST,
\label{eq:2.17}
\eea
where ST denotes the surface term which arises on converting
$-\partial_{z'}\delta (z-z')$ to
$+\partial_{z}\delta (z-z')$ and integrating by parts with
respect
to z.
\bea
     ST &=& = \int dz' \delta \ut{N}(z') \cdots \{\E{a}{C}(z')[
\delta (z-z')
               \A{D}{b}(z)]_{z=z_l}^{z=z_r}\}
                    \nonumber \\
            &=& 0.
\label{eq:2.18a}
\eea
At a first glance one might think that surface terms vanish only
when integrating by parts with
respect to z', since the $\delta \ut{N}$ depends on z', not z.
Because of the $\delta (z-z')$,
however,  $\delta \ut{N}(z')$ kills surface terms arising from
integration by parts with respect to
either varable, z or z'. $\Box$

     The discussion just given proves that the first constraint
$\int \E{a}{B}\A{B}{b}$
in \eq{2.9b} is physical, but says
nothing about its complex conjugate partner $\int  \E{a}{B}
[-\A{B}{b} +  2 Re \A{B}{b}] $.
In the previous section and in part B of the present section I
remarked that the BPR
constraints come in complex conjugate pairs, for example,
   \bea
     \A{-}{a} &=& 0; \nonumber \\
      -\A{+}{a} + 2 Re \A{+}{a}  &=& 0.
\label{eq:2.18}
\eea
The second member of the pair is much harder to work with,
because
of the non-polynomiality
of the $Re \A{B}{a}$ factor.   This , plus the complex conjugate
pairing,   suggests that an
indirect approach might save labor: do the calculation for the
first constraint, then  take the
Hermitean conjugate to obtain the
corresponding result for the harder constraint.  For example,
commute the first constraint in \eq{2.9b} with
$\hat{C}_i$, then take
the Hermitean conjugate of the commutator to obtain the
commutator
of the
second constraint with  $\hat{C}_i$ .  This indirect approach
works
when $\hat{C}_i$ is the Gauss
or diffeomorphism constraint,  because these two constraints are
self-adjoint.
However,  the scalar constraint is not self-adjoint for my factor
ordering, or indeed, for any of
the factor orderings  most often encountered in practice when
working with a complex connection.  Therefore one cannot simply
take a Hermitean conjugate to prove that the second constraint in
\eq{2.9b} is physical; one must directly work out the commutator
of
the second constraint with $\hat{C}_S$.  I now do this.

     Proof that the second constraint of \eq{2.9b} is physical:
since the sum of the two constraints of \eq{2.9b} is proportional
to  \E{b}{B} Re \A{B}{a}, one need only  prove that this
expression
(or
more precisely its integral over  z) commutes with the scalar
constraint.  The  Re \A{B}{a} factor  is non-polynomial; but
fortunately the
calculation is not involved, because \E{b}{B} Re \A{B}{a} turns
out
to be a total derivative.  Therefore its integral is a surface
term, and the surface term is physical, because
$\hat{C}_S$ is smeared by a $\delta \ut{N}$ which vanishes at the
surfaces.  The following formulas may be used to prove that
\E{b}{B} Re \A{B}{a} is a total derivative:
\bea
           Re  \A{B}{a}&=&\epsilon _{BJ} \omega^{ZJ}_a ;\nn
     \omega ^{IJ}_a &=& [\Omega _{i[ja]} + \Omega _{j[ai]}
                         - \Omega_{a[ij]}]e^{iI} e^{jJ};\nn
     \Omega _{i[ja]} &=& e_{iM}[\partial _je^M_a
                                   - \partial_ae^M_j]/2.
\label{eq:2.19}
\eea
The first line is the reality condition for the transverse
connections  \A{B}{a}.   The Lorentz
connection $\omega ^{IJ}_a $ is four-dimensional; $e_{iM}$ is the
tetrad, and $
e^M_i $ is its inverse.   The standard Lorentz gauge fixing
condition, $e^{tZ} = 0$, may be used
to simplify the sums over i in the definition of $\omega ^{IJ}_a
$,
when I = Z; one gets
\bea
      \omega ^{ZJ}_i &=& -\partial _zg_{ij}e^{zZ}e^{jJ}/2; \nn
     \E{b}{B} Re \A{B}{a} &=& e e^b_B \epsilon _{BJ}
                    \omega^{ZJ}_a \nn
                    &=& -\partial _zg_{aj} \epsilon^{bj}/2,
\label{eq:2.20}
\eea
which is a total derivative as required. $\Box$

     It is fortunate that both constraints in \eq{2.9b} are
physical; otherwise the scalar constraint
would  treat the two
polarizations of gravitational waves
differently, since the two constraints  are needed in
\eq{2.9b} because there are
two polarizations.   It would not be surprising if the two
polarizations are treated differently in
the wavefunctional, since that quantity is not directly
observable.
In fact the two polarizations
{\it are} treated differently, at least in the wavefunctional for
the linearized limit \cite{linpsi}.  However, it would be
surprising if the two polarizations have different
behaviour under commutation with
the scalar constraint.  That would lead one to question the usual
choices of factor ordering, since
this ordering causes the scalar constraint to be non-Hermitean, and
a Hermitean constrain
would treat the two polarizations symmetrically.  .
In section VI, I discuss indirect
evidence that the factor ordering should be changed; but
I have no direct evidence that
the ordering is incorrect.

     (This paragraph probably can be skipped on a first reading.)
In the plane wave case one has
special (z,t) coordinates, call
them (z',t'), which
parameterize the hypersurfaces $z' \pm ct' = {\mbox const}$.  One
can therefore envisage
solving the problem of meaningless z coordinate in another way,
by
a transformation of
coordinates  to the more meaningful coordinate z'.   One might
still integrate over dz, but with a
factor of $\partial z'/\partial z$ inserted in the integrand.
(This factor would give the desired
density weight of one.)    I have perhaps not given this option
the
attention it deserves.  Note,
however, it would require a {\it great deal} of attention.
Information about the special surfaces
z' = const. is encoded in the metric, so that the factor
$\partial
z'/\partial z$ is not just a function,
but a functional of the metric components ( and a highly
non-polynomial function at that!)
This factor becomes an operator in the quantum case, and it is a
highly non-trivial question how
one regulates and factor orders this operator.  This is the
problem
one runs into if one shifts to z'
at the quantum level.  If one trys to shift at the classical
level,
then one must fix the $C_z$ gauge
and replace  Poisson brackets by Dirac brackets which respect the
gauge-fixing condition.  The
classical Dirac brackets between the surviving observables tend
to
be messy, and so far I have
found no quantum representation.

     (The following three paragraphs contain material which at
first glance may seem to be of
only historical interest,  but  will be needed later
in section V.)  I interpret the BPR
criteria in a semiclassical
sense, as $<\hat{F}> = 0$ rather
than $\hat{F} \psi = 0$.  Yet the Hamiltonian constraints are
always
imposed strongly, as $\hat{C}_i
\psi = 0$, even though (in the linearized limit, at least) the
$\hat{C}_i$ are sums of creation and
destruction operators, like the BPR field strengths.  Why
this difference in treatment?
This same question was posed and answered in a different context,
Lorentz gauge QED, many
years ago, and it is worthwhile to take a moment here to review
that discussion \cite{Lornorm}.  In  Lorentz
gauge QED, the analog of the $C_i$ is (the usual Gauss
constraint,
plus) the four-divergence
$\partial A = 0$.   The analog of the
strong requirement  $\hat{C}_i \psi = 0$ would be  $\partial
\hat{A} \psi =
0$; and the analog of the
semiclassical requirement $< \hat{C}_i > = 0$ would be $\partial
\hat{A}
^{+}\psi = 0$, where the superscript +
denotes positive frequency components.  (Since a splitup into
positive and negative frequencies
is available in QED, there is no need to introduce a
semiclassical
average.)

     Both constraints,
the strong and the positive frequency/semiclassical, are used in
the Lorentz gauge literature.
Authors who employ the positive frequency constraint tend to treat
the ``unphysical''   part of the
Hilbert space with more respect.  (Remember that the Lorentz
gauge
condition is designed to
eliminate the effects of the unphysical, longitudinal and
timelike
components.)   Heitler \cite{Heitler} is a
typical proponent of this approach: he gives a very careful
treatment of the
unphysical sector, including a  full
discussion of the Gupta-Bleuler formalism.  The payoff is that
dot products over the full Hilbert
space are well-defined, including dot products of longitudinal
and timelike photons.  Authors
who employ the stronger constraint \cite{Dirlec} pay a price: it is
possible to
find states which are annihilated
by both the annihilation {\it and} the creation parts of
$\partial
\hat{A}$, but these states are not
normalizable in the unphysical sector \cite{Lornorm}.  This result
is not particularly
surprising: since
the creation operators in $\partial
\hat{A}$ create a state with one more timelike or longitudinal
photon,
$\psi$ must be a sum over an
unbounded, infinite number of longitudinal and timelike
occupation
numbers.  This infinte sum
leads to the divergence in the norm.   The authors who use the
strong constraint are well aware
of this difficulty, and they circumvent it by requiring that the
dot product  in Hilbert space be
taken over physical excitations only.

      Returning to the gravitational case, one can now see why
the strong criterion will work
for the $C_i$, but not for the BPR operators.  For the
moment, imagine the gravitational
theory to be linearized, so that the analogy to QED is strongest.
The creation operators for the
BPR operators create physical quanta, not unphysical.  If
I impose the strong criterion,
I get states which have an infinite norm in the physical sector.
There is no way of avoiding
this by restricting the measure at a later step.  If I
now pass from the linearized to the
full theory, there is no reason why the strong criterion should
suddenly become applicable.  I must use
the semiclassical criterion, which is justified using the
correspondence principle.

\section{Additional Solutions}

     In the previous section I derived quantum BPR operators.  In
the
present section I construct  new solutions to the
constraints.
 In the next section I apply the BPR, ADM energy, and $L_Z$
operators to the solutions
constructed in paper II-III and this section.

     I start from the  solutions considered in
III.  These are strings of
transverse \E{a}{A} operators, ordered along the z axis, and
separated by holonomies:
\be
     \psi = [\prod_{i=1}^n \int_{z0}^{z_{n+1}} dz_i
\Theta(z_{i+1}
-z_i) M(z_{i+1},z_i)
               \E{a_i}{A_i}(z_i) S_{Ai} \Theta(z_{1} -z_0)]
M(z_{0},z_{n+1}).
\label{eq:4.1}
\ee
The M are holonomies along z,
\be
     M(z_{i+1},z_i) = exp[i\int_{z_i}^{z_{n+1}}\A{Z}{z}(z') S_Z
dz'],
\label{eq:4.2}
\ee
and the $S_M$ are the usual Hermitean SU(2) generators.  These
can
be 2j+1 dimensional; they
need not be Pauli matrices.   The $\Theta $ functions in \eq{4.1}
are Heaviside step functions which
path-order the integrations, $z_0 \leq z_1 \leq \cdots \leq
z_{n+1}$.  For this
section only, the boundary points $z_l$ and $z_r$ are relabeled
$z_0$ and $z_{n+1}$ .  Although the metric is not flat at the
boundaries,
it can be taken as conformally
flat at boundaries, with any radiation
present  confined to a wave packet  near the origin \cite{II}.

     Since the full SU(2)
invariance has been gauge fixed to O(2),
it is convenient to use basis fields introduced at \eq{1.3},
fields which are  irreducible
representations of O(2).  These are
one-dimensional, labeled by the eigenvalue of $S_Z$, e.\ g.\ ,
\bea
      \E{a}{\pm}& =& (\E{a}{X} \pm i\E{a}{Y})/\sqrt{2}; \nonumber
\\
     \E{a}{A} S_A &=& \E{a}{+} S_- \mbox{ or }   \E{a}{-} S_+ .
\label{eq:4.3}
\eea
Because the irreducible representations are one-dimensional,
there
is no need to sum over both
values of $A_i = \pm$ in \eq{4.1}, in order to obtain a
Gauss-invariant expression; nor is it
necessary to take the trace in that equation.  However, one must
be
sure to have an equal number
of $S_+$ and $S_-$ matrices in the chain, in order to form a
closed
loop of flux with no open
ends violating Gauss invariance.  That is, if one visualizes each
holonomy $M(z_{i+1},z_i)$ as a
flux line along z from $z_i$ to $z_{i+1}$, then the factor in the
square bracket, \eq{4.1},  may
be
visualized as a flux line from $z_0$ to $z_{n+1}$.  The line varies
in thickness (varies in $S_Z$
eigenvalue) because of the $S_{\pm}$ operators encountered along
the way, but the final $S_Z$
value at $z_{n+1}$ must equal the initial $S_Z$ value at $z_{0}$
(there must be an equal
number of $S_+$ and $S_-$ matrices in the chain).  Then the final
holonomy in \eq{4.1},
M($z_{0}$,$z_{n+1}$), can join the two  ends at $z_0$ and $z_{n+1}$
and
turn the open flux line
into a closed flux loop.   As shown in III, the
wavefunctional of \eq{4.1} can be made to satisfy all the
constraints, by suitable choice of the
$a_i$ and $A_i$.

     In order to obtain a larger, kinematical space, as well as
more physical solutions, one can
simplify (!) \eq{4.1} by dropping all $\Theta$ functions.  This
removes the path ordering, or (
in visual terms) this  allows flux lines to double back on
themselves.
\be
     \psi_{kin } =  \prod_{i=1}^n \int_{z0}^{z_{n+1}} dz_i
M(z_{i+1},z_i)
                    \E{a_i}{A_i}(z_i) S_{Ai}  M(z_{0},z_{n+1}).
\label{eq:4.4}
\ee
Again, the expression is Gauss invariant even if $A_i = \pm$ is
not
summed over; and no trace
is needed.

     To check the spatial diffeomorphism and scalar
constraints, one must first obtain
these constraints
from  the Hamiltonian, written out in  an O(2) eigenbasis:
\begin{eqnarray}
     H_T &=& \Np [i\Etwo (\E{z}{Z})^{-1} \epsilon _{ab}\A{+}{a}
\A{-}{b}
     +\sum_{\pm} (\pm i) \E{b}{\pm} {\rm F}^{\mp}_{zb}]
\nonumber
\\
     & & + iN ^z \sum_{\pm} \E{b}{\pm} {\rm F}^{\mp}_{zb}
\nonumber
\\
          & & -i{\rm N_G} [\partial _z\E{z}{Z} - \sum_{\pm} (\pm
i)
                     \E{a}{\pm} \A{\mp}{a} ] + ST \nonumber \\
     &\equiv & \Np C_S + N^z C_z + {\rm N_G}C_G +   ST,
\label{eq:4.5}
\end{eqnarray}
where
\begin{equation}
     {\rm F}^{\mp}_{zb} = [\partial _z \mp i\A{Z}{z}] \A{\mp}{b}.
\label{eq:4.6}
\end{equation}
\Etwo\  is the determinant of the 2x2 transverse subblock of the
matrix \E{a}{A}.    ST denotes
surface   terms
(terms evaluated at the two endpoints on the z axis, $z_0$ and
$z_{n+1}$).  The
detailed form of these terms is
worked out in II but will not be needed here.  The primed lapse
\Np\ equals  the usual lapse N
multiplied by a  factor of \E{z}{Z}, and correspondingly the
scalar
constraint $C_S$ is the
usual constraint divided by \E{z}{Z}.  As shown in II, this
renormalization leads to a much
simpler constraint algebra, but again the details of this will
not
be relevant here. The system is
quantized by replacing transverse \A{A}{a} and \E{z}{Z} by
functional derivatives, these being
the fields conjugate to the fields in $\psi$.
\bea
     \A{\pm}{a} &\rta & \hbar \delta / \delta \E{a}{\mp};
\nonumber
\\
     \E{z}{Z} & \rta& - \hbar \delta / \delta \A{Z}{z}.
\label{eq:4.7}
\eea
The operator ordering (already adopted in \eq{4.1}) is functional
derivatives to the right.  The
first term in \eq{4.5} contains an inverse  operator
$(\E{z}{Z})^{-1}$; this is well-defined
provided  \E{z}{Z}M  never vanishes, that is, provided the $S_Z$
in \eq{4.2} never has the
eigenvalue zero.

     The physics must be invariant under small {\it changes}
$\delta N$ in the lapse and shift,
so that the constraint should be written as
\bea
     0 &=& \int dz [\delta \Np C_S +\delta N^z C_z ]\psi_{kin};
                    \label{eq:4.8}  \\
     0 &=& \delta \Np (z_b) = \delta N^z (z_b),
\label{eq:4.9}
\eea
where $z_b$ is either boundary point, $z_0$ or $z_{n+1}$.
\Eq{4.9} guarantees that the
boundary conditions at $z_b$ are left  unchanged by the
transformation of  \eq{4.8}.

     Both the $C_S$ and $C_z$ constraints  in \eq{4.8} contain
terms proportional to $ {\rm
F}^{\mp}_{zb} $,
the field strength defined at \eq{4.6}.  When a typical term of
this type  acts on $\psi$, the
result is (up to constants)
\bea
     \int dz \delta N(z)\E{a}{+} {\rm F}^{-}_{zb}
          [\psi_{kin}]&& \nn
      =&&\int dz \delta N(z) \E{a}{+} {\rm F}^{-}_{zb}
          [\cdots \int dz_i M \E{a_i}{+} S_- M \cdots ] \nn
      =&&\int dz \delta N  \sum_i  \E{a_i}{+}[\cdots
               \int dz_i   M (\partial_z
                  - i\A{Z}{z}) \delta (z - z_i) S_- M \cdots] \nn
      =&&\cdots [\cdots  \int dz_i \delta (z - z_i)
               (\partial _{zi}-
                         i\A{Z}{z})M S_- M \cdots ] \nn
       =&&\cdots [ \cdots  \int dz_i \delta (z - z_i)
               (-iM [S_Z,S_-]M \A{Z}{z}
                         -i \A{Z}{z}M S_- M) \cdots ] \nn
       =&&0.
\label{eq:4.10}
\eea
On the third line I have changed the $\partial _z $ to $\partial
_{zi} $ and integrated by parts with
respect to $z_i$.  The surface terms at $z_i = z_b$ vanish
because
the $\delta N (z) \delta (z -
z_i)$ yields a factor of $ \delta N (z_b) $, which vanishes  at
boundaries.  Thus $\psi_{kin}$ is
annihilated by all constraint terms  containing  field strengths
${\rm
F}^{\mp}_{zb} $.   This is already enough  to prove that the
spatial diffeomorphism constraint
$C_z$ annihilates $\psi_{kin}$, hence  $\psi_{kin}$ is at least
part of a
kinematical basis, if not a physical basis.

     The state would be physical if it were also annihilated by
the
first term in $C_S$, which I
call $C_E$ because of the \Etwo\  factor which it contains.  (This
term also happens to be proportional to a
field strength, namely $F^Z_{xy}$.)   The following is a
sufficient condition for $\psi_{kin}$ to
be physical: $C_E$ annihilates the state if it
contains only two out of the four  transverse fields:
\bea
     \mbox{(either)} & & \E{x}{+} \mbox{\ and\ } \E{x}{-}
          \mbox{\ only\ }; \nonumber \\
     \mbox{(or)} & & \E{y}{+} \mbox{\ and\ } \E{y}{-} \mbox{ only}.
\label{eq:4.11}
\eea
If $\psi $ contains  \E{x}{+}  and \E{x}{-} only, for instance,
their product \E{x}{+} \E{x}{-}
is not contained in the triad determinant \Etwo; therefore the
connection determinant $ \epsilon
_{ab}\A{+}{a} \A{-}{b}$ in $C_E$ will necessarily annihilate any
wavefunctional containing
only \E{x}{+}  and \E{x}{-}.  Note the wavefunctional must have
an
{\it equal} number of
\E{x}{+} and \E{x}{-}fields, because Gauss invariance requires an
equal number of $S_-$ and
$S_+$ operators in the chain.

     One can generate additional physical states, starting from
those described by \eq{4.4}
and \eq{4.11}, by applying the  operators $G^x_y $ or $G^y_x$,
where the $G^a_b$ are integrals
over z of the weight one objects constructed at \eq{2.9b}:
\be
     G^a_b = \int _{z0}^{z_{n+1}} dz \E{a}{A} \A{A}{b}.
\label{eq:4.12}
\ee
As shown in section III, the operators $G^a_b$ commute with
H; they are physical. Hence application of m factors of
$G^y_x$ to a functional $\psi [
\E{x}{+}, \E{x}{-}]$ replaces m x superscripts in the chain by y
superscripts, but leaves $\psi$
physical.  The operators $G^a_b$ change the total intrinsic spin
of the wavefunctional; for a discussion of this point, see
section VI.

     I have labeled the generators $S_{Ai} $ in \eq{4.1} using
two quantum numbers, j and
m (more precisely, initial or final m, if the generator $S_{A}
$is one which changes m).  Once
the SU(2) symmetry is broken to O(2), however, the j quantum
number loses signifigance.  I
could replace the $S_{Ai} $ by any other matrix with the same m
(and $\Delta$m) but different
j, and $\psi$ would change only by a constant factor.

     Even though the j has no physical signifigance, it is
mathematically convenient to use
$S_A$ having definite j.  One can then employ the familiar
commutation relations of the $S_A$
in calculations.  Also, the planar state $\psi$ is presumably a
limit of some three-dimensional
state for which the label j has meaning.  The fact that states of
different j are equivalent in the
planar limit presumably means that the correspondence between
three-dimensional and planar
states is many to one.

\section{Application to Solutions}

       In this section I study the solutions constructed so far by
applying several operators to  them:
the modified BPR operators (from section III),   the ADM energy
operator (from paper III and
appendix C), the area operator \E{z}{Z} for areas in the xy plane
(from paper III), and the
operator $L_Z$ giving the total spin angular momentum around the z
axis (from paper IV and appendix B).
All the  solutions (those constructed in papers II-III as well as
the new solutions constructed in
section IV)  have
the form of  strings of transverse $\E{a}{\pm}(z_i) S_{\mp} $
operators
separated by holonomies $M(z_{i+1},z_i)$ .  The solutions in papers
II-III have additional  step
functions $\theta(z_{i+1}-z_i)$  which path order the integrations
over the $z_i$ , but the
$\theta$  factors will  play a minor role in the considerations of
the present section.

     Consider first the ADM energy operator.  Often this is
identified with the surface term in
the Hamiltonian, but, as discussed in appendix C, the volume term
can also contribute.  In the
present case the volume term typically does  contribute, but its
only effect is to double the size of
the surface term, and I will ignore volume contributions.  The
surface term is, from \eq{d.4},
\bea
      H_{st} &=&-\epsilon_{MN}\E{b}{M}\A{N}{b} \mid_{z_l}^{z_r}
                                               \nonumber \\
           &=& i\hbar[\E{b}{-}\delta /\delta \E{b}{-}
                - \E{b}{+}\delta /\delta \E{b}{+}]_{z_l}^{z_r}
\label{eq:5.1}
\eea
When this operator acts upon a factor of $ \E{a}{\pm}(z_i) dz_i$ in
the wavefunctional, it gives
$\mp i \hbar \E{a}{\pm}dz_i$ times a factor of $\delta (z_i -
z_r) $ or $\delta (z_i - z_l)$.
Obviously none of the solutions is an eigenfunction of the ADM
energy, since the $\delta$
function deletes one integration $dz_i$.  One could perhaps
construct an eigenfunction by
summing over an infinite number of solutions, each containing one
more $dz_i$ integration.
Each additional integration should be multiplied by an additional
factor of i, to cancel the i in
\eq{5.1} and make the eigenvalue real.  Investigation of such sums
is beyond the scope of the present paper.  Without a measure one
does not know whether such a sum converges to a normalizable
result.

     If the Gauss constraint,
\be
     C_G = -i[\partial_z\E{z}{Z} -\epsilon_{MN} \E{a}{M} \A{N}{a}],
\label{eq:5.1a}
\ee
vanishes at the boundaries, it can be used  to simplify the
ADM surface term to
\bea
     H_{st}& =& -\E{z}{Z},_z \mid_{z_l}^{z_r} \nn
          &=& \hbar (\delta /\delta \A{Z}{z} ),_z.
\label{eq:5.2}
\eea
Unwanted factors of i are now more of a problem.  \A{Z}{z} occurs
only in holonomies, where
it is always multiplied by (a real matrix $S_Z$ times) a factor of
i, and the \E{z}{Z} will bring down this factor of i.
Except for the solutions
constructed in paper II, there is always at least one boundary
where the Gauss constraint holds, so
that factors of i will be a generic problem.  Of course one could
eliminate the problem by discarding the holonomy structure, but
this is a solution almost as unattractive as the problem.

     The operator $\E{z}{Z},_z$ occurs in  the ADM energy, while
\E{z}{Z} itself is the area
operator for areas in the xy plane.  (\E{z}{Z} = e$e^z_Z$ =
$^{(2)}e$.)  Therefore the area operator
also has pure imaginary
eigenvalues, a situation already noted in Appendix D of paper II;
see also DePietri and Rovelli \cite{DePRov}.

     Even if one for the moment ignores the factors of i, there is
another problem with the
area operator: at any boundary where the Gauss constraint is
satisfied, the area operator will
always give zero.  If the Gauss constraint is satisfied, say at the
left boundary $z_l$, then there is no net flux
exiting at $z_l$.  One can regroup the holonomies  until there are
no M($z_i,~z_l$), only M($z_r,~z_i$); or until every M($z_i,~
z_l$) is paired with an M($z_l,~z_j$) to give  M($z_i,~z_l$)
M($z_l,~z_j$) =  M($z_i,~z_j$).   Either way, there is no holonomy
depending on
\A{Z}{z}($z_l$), and the area operator \E{z}{Z}($z_l$) gives zero.

     Next consider the action of the BPR operators.  The solutions
given in II
(and some of those considered in
section IV) contain either  \E{a}{+} operators or  \E{a}{-}
operators, but
not both.  A wavefunctional which contains only \E{a}{-} operators
(for example) will be
annihilated by $\A{-}{a} = \hbar \delta / \delta \E{a}{+}$, even
before any semiclassical
average is taken:
\be
     \A{-}{a}\psi [\E{a}{-}] = 0.
\label {eq:5.3}
\ee
In classical theory, \A{-}{a} = 0 is a signal that the solution is
pure left-moving; and from this one might expect that $\psi
[\E{a}{-}] $ is unidirectional.
Condition \eq{5.3} is too strong, however.  As discussed at
\eq{2.3} of section III,
One expects at most a vanishing semiclassical average $\langle
\A{-}{a}\rangle$ = 0.   In fact from the
remarks on Lorentz gauge QED in the concluding paragraphs of
section III, \eq{5.3} implies that $\psi$ is probably not a
normalizable state!

     The solutions of paper III and most of those from section IV
contain both  \E{a}{+} and
\E{a}{-}, hence are not annihilated by any BPR operator.  Onc
cannot conclude that these
solutions  are infinite norm, therefore. However, they do suffer
from the problems described previously in this section, those
associated with the ADM energy and area operators.

\section{Conclusions}

     Papers II-III proposed new solutions to the constraints, and
section IV of the present
paper proposes still more solutions.  However, the
investigations of section V have demonstrated that  these
solutions
are less than satisfying in
several respects.

     This outcome is perhaps not surprising.
In  earlier work , simplicity was the primary criterion for
choosing the method
of  quantization (polarization and factor
ordering) ,  as well as  the primary criterion for constructing
solutions.   Simplicity is the criterion one uses when
information is scarce, however. In the present paper several
operators of  physical signifigance
were available, and  the theory and its solutions can be held to a
standard more demanding than simplicity, the standard of
a reasonable physical interpretation.  As a result, solutions
are called into doubt; but this happens for a reason which is
fundamentally positive: more is known about how to interpret and
understand the solutions.

     The  difficulties with  imaginary eigenvalues encountered in
section V  seem to require fundamental revisions in the theory,
since the difficulties are closely linked to
the complex nature of the
connection.   As discussed in section V, the operator \E{z}{Z},
present in both the area
operator and the ADM energy
operator,  has complex eigenvalues because of the i in the
holonomies, exp[i\A{Z}{z}$\cdots$].  Getting rid of the i entails
dropping or modifying the holonomic structure, not a pleasant
prospect.   Recently
Thiemann \cite{Thie} has proposed an alternative formalism based
upon a real
connection.    Thiemann's alternative is motivated primarily by
issues of regularization, but has the desirable side effect of
producing real eigenvalues for the area operator.

     It is a little harder to see how switching to a real
connection will cure the problem of zero
eigenvalues of the area operator at boundaries, also uncovered in
section V.  The zero
eigenvalues occur only at boundaries where Gauss invariance is
satisfied .   Since \E{z}{Z} is a Gauss
invariant, at first sight any
connection between area and Gauss invariance seems strange.  The
connection is indirect, via
the structure of the complex connection \A{Z}{z} = i Im~\A{Z}{z}
+ Re~\A{Z}{z}.    Im~\A{Z}{z} is the part which does not commute
with \E{z}{Z}; therefore its presence in the
wavefunctional gives rise to non zero area.  Re~ \A{Z}{z} is the
part which transforms like a
connection under Gauss rotations; therefore it is needed in the
wavefunctional for gauge
invariance.   The notions of area and gauge invariance are linked
only because Im~A and Re~A
are linked together to form a single (complex) connection.  At a
boundary where Gauss invariance is satisfied, if  there is no net
flux
exiting, then there is no dependence on the connection, hence no
area.
In the Thiemann scheme one still joins
Im~ A and Re~ A together to form a single (real) connection; in
fact the Thiemann connection is just
the Ashtekar connection without the factor of  i.  The real and
imaginary parts of the connection are separated when
constructing the Thiemann constraints.  Must they
be separated when constructing the wavefunctional as well?
If the answer is yes,
the wavefunctional would not be purely a product of holonomies.

   Before doing anything as drastic as dropping the holonomic
structure, it is a good idea to investigate what happens to the
zero area argument when it is extrapolated from the planar case
to the full, three-space-dimensional
case.  The argument that Gauss invariance leads to zero area
depends on properties of the
wavefunctional
at boundaries, and the behaviour at boundaries changes markedly
with spatial dimension.

     In the full three-dimensional case, the
smearing function for the Gauss constraint must vanish at spatial
infinity, so that there is no need for the Gauss constraint to
annihilate the wavefunctional there.  As a result, net flux  may
pass through the boundary at infinity, and there is no difficulty
obtaining finite area at the boundary, even when the
wavefunctional is purely a product of holonomies.   This
suggests that the zero area  may be  a
problem which occurs only in the planar limit.

     In fact the problem may not exist even in the planar limit,
if one takes this limit correctly.   Imagine the generic three
dimensional flux configuration which is well approximated by
planar symmetry: near the origin, the flux lines corresponding to
holonomies containing
\A{Z}{a} $dx^a$ are finite in cross section and well-collimated
along the z axis.   The planar wavefunctionals constructed in
papers II-III contain factors of
$S_{\pm}$, presumably relics of Clebsch-Gordan coefficients
coupling \A{Z}{a} $dx^a$ holonomies to holonomies containing
\A{X}{a} $dx^a$ and \A{Y}{a} $dx^a$.  The latter are represented
by flux lines lying in planes z = constant.  Although \A{Z}{a}
$dx^a$ flux lines are well collimated near the origin, they must
diverge far out along z into the past or future.  A sketch of the
\A{Z}{a} $dx^a$ flux lines
resembles a  drawing
depicting radial geodesics near a wormhole:  the flux lines  come
in from radial
infinity, pass through a narrow ``throat''   oriented along z,
then
diverge once more to radial infinity.  (The wavefronts
perpendicular to these rays are constructed from \A{X}{a} $dx^a$
and \A{Y}{a} $dx^a$ holonomies.)  Alternatively,  the \A{Z}{a}
$dx^a$ flux
lines at infinity may not exit through the surface at infinity, but
may loop back and close on themselves, resembling the flux lines
of a solenoid in magnetostatics.  Either outcome is allowed by
the boundary conditions on the Gauss smearing function at
infinity, and
for either behavior at infinity, the behavior at the throat is
the same.  If one takes a cross section through two points $z_l$
and $z_r > z_l$ at the throat, one finds net \A{Z}{a} $dx^a$ flux
through both
boundary points $z_l$ and $z_r$.  If this picture of the three
dimensional flux
is correct, then in the planar limit one should {\it not} impose
Gauss invariance at the boundaries.  Planar solutions would
resemble the
``open flux''  solutions studied in paper II.  If there is
\A{Z}{a} $dx^a$ flux
through the boundaries, the zero area problem at boundaries
disappears.

     Even though the \A{Z}{a} $dx^a$ flux lines now extend
throughout the entire
range $z_l \leq z \leq z_r$, one can still construct localized
wave packets.  In paper II I reviewed the geometrodynamical
treatment of the planar problem, and introduced the Szekeres
scalar fields B, W, and A.  (It is a little easier to work out the
boundary conditions for B, W, and
A, rather than work directly with the Ashtekar fields; as shown in
II, the boundary conditions  on B, W, and A then
imply corresponding  boundary conditions on the Ashtekar
variables.)   When only the field A is
present, the
forces on a cloud of test particles are isotropic in the
transverse (xy) direction; the elliptical distortions
characteristic of gravitational waves appear only when the B and
W fields are non-zero.  (In a more covariant language, the
components of the Weyl tensor which
give rise to  transverse deviations of geodesics are present only
when B and W are non-zero.)
One can impose wave packet boundary
conditions on B and W (equivalently,  on transverse components of
the Weyl tensor), requiring
these quantities to vanish at
the boundaries $z_b$; but  this requirement  tells us nothing about
the behavior of A  at the
boundaries.  The variable A determines geometrical quantities such
as areas: the Ashtekar area operator \E{z}{Z} is just exp(A).
Perhaps one has a ``wave
packet'',  but not a ``geometry packet''.   One may
require localized, wave packet behavior for B and W, but not for
the more geometrical quantity A.

     To summarize, there are two possible solutions to the zero
area problem.  The first splits the connection, abandoning the
strict holonomic form for the wavefunctional.  The second
allows  the Gauss constraint to be non-zero at boundaries  and in
effect assumes that the
``open flux'' boundary conditions used in paper II are generic.
Further information and thought is meeded before one can decide
between these two  alternatives.

     Apart from the zero area difficulty, there is another reason
why the transition from three to one space dimension needs
more attention.  Ultimately one would like to use the planar
case as a guide to the behavior of radiation in the full, three
dimensional case.  In the full case, one expects the connections
to occur in holonomies, so as to
preserve gauge invariance.  In the planar case, the gauge fixing
allows the transverse connections
to occur outside of holonomies.   Two key radiative  properties
of the planar solutions, their
directionality and spin, are associated with the transverse
sector, which  least resembles the
three-dimensional case.  It will probably be necessary to recast
the transverse sector in a more holonomic
language, in order to understand more clearly what happens on
passing to the full theory.

  For the moment let us overlook any possible difficulties with
zero area and suppose that one
shifts to a real connection, in order to eliminate the problem
with imaginary eigenvalues.   One can ask whether the
solutions constructed in papers II-III and section IV are likely
to survive the shift to a real connection formalism.   The
present solutions may not survive, if the
factor ordering is changed; and it is easy to imagine a reason
why one might want to change the factor ordering.  The scalar
constraint usually must be taken to be non-Hermitean, in a
complex
connection formalism; whereas with a real connection one may wish
to factor order so as to make the constraint Hermitean.

          It may be helpful to comment briefly on why the scalar
constraint is difficult to make
Hermitean in a complex connection formalism.  The usual recipe
for constructing a self-adjoint
operator is to factor order it, and
then, if the operator is not self-adjoint, form the average $(C +
C\dagger )/2$.  $C \dagger$ is C,
with the order of all operators reversed and  the connections A
replaced by $A \dagger$; in turn
the $A \dagger$ are replaced by -A + 2 Re A.    This last step
introduces the unwanted
non-polynomial expressions Re A into the $C\dagger$ term.   In the
case of the Gauss and spatial
diffeomorphism constraints, identities may be used to eliminate
the Re A contributions, and $C\dagger$ is  well-behaved, in fact
identical to C.  In the case
of the scalar constraint, the
unwanted Re A terms do not go away.   Within a real connection
framework, the  $A \dagger$ is
just A, and the traditional $(C + C\dagger )/2$ recipe is easier
to implement.

     Although the present solutions may not survive as
exact solutions, they may
constitute approximate solutions to the new,  modified
constraints,
solutions valid in the limit $ \hbar \rta 0$.  This would happen
because the new and old scalar
constraints presumably will differ only by a reordering of
factors, hence will differ by terms of
order $\hbar $.   Also, the $G^a_ b$ operators, defined at
\eq{2.10b}, wer shown to be constants
of the motion by Husain and Smolin \cite{HSm}.  These operators are
likely to remain constants
of the motion, in any transition to a new operator ordering,
because of the close connection between the  $G^a_ b$ and total
spin  \cite{IV}.

     Even though the complex connection formalism may not be
appropriate for the dynamics, this
formalism is the natural one to use when constructing a criterion
for the presence of radiation.
Note that the BPR operators were derived in section II using only
symmetry considerations;
no assumptions were made about the factor ordering or dynamics.
Even if one dropped the
Ashtekar connection and used the Thiemann connection,
one would have to reintroduce the Ashtekar connection in order to
express the results of section
II succinctly!   Anyone familiar with the classical results on
radiative criteria will not be
surprised at this: much of that work is most conveniently
expressed using the language of
complex connections.   (See for example the work on the Weyl
tensor quoted in section I and Appendix D.)

     My original intention in constructing the BPR operators was
to elucidate the spin and directional character (that is, left
vs.\
right-moving character) of radiative quantum states.  Since these
operators turned out to be semiclassical in character, and no
measure is yet available, it is perhaps early days to make a
final judjment on the efficacy of the BPR operators.   However,
it
is probably true that any
unidirectional operator (BPR or other) will be at most
kinematical; that is, it will commute with
the Gauss and diffeomorphism constraints, but not the scalar
constraint.   Conversely, any
operator which is physical (commutes with all constraints) will
not be unidirectional; i.\ e.\ it
will mix BPR operators of opposite directionality.  The concept
of unidirectionality is well
defined in classical theory; see section II and appendix B.  In
quantum theory, however, every
unidirectional wave must traverse a vacuum filled with zero point
fluctuations which are moving
in both directions.  Presumably this is the intuitive reason why
the modified BPR operators constructed in section III are either
unidirectional or physical, but not both.
Consistent with this interpretation of unidirectionality as
primarily a classical concept, it is
possible to prove that the
unidirectional BPR operators are physical in {\it classical}
theory; but factor ordering problems
prevent the proof from going through in the quantum case.
Presumably any criterion for
unidirectionality will have to be at most a
semiclassical one.

     Classically, exact plane wave solutions are known in which the
area operator \E{z}{Z} = $^{(2)}e$ evolves to zero \cite{Szmet}.
In
fact this collapse behavior appears to be generic; solutions which
do not collapse are rare and are unstable under small perturbations
\cite{Yurt}.  The zero area cannot be removed by a change of
coordinates, since typically there is an accompanying singularity
in a scalar polynomial quadratic in components of the Weyl
curvature tensor.  One can ask whether a quantum-mechanical effect
might prevent this collapse.  It is not possible to answer this
question definitively within the present context, because the area
operator has imaginary eigenvalues.  Nevertheless, one can see the
outlines of a possible quantum solution which would avoid a
collapse.  The quantum area operator \E{z}{Z} acts on holonomies
exp(i$\int$ \A{Z}{z} $S_Z$ dz); so long as $S_Z$ is not allowed to
assume the value zero, \E{z}{Z} cannot have the eigenvalue zero.
In the solutions constructed in papers II-III and section IV, the
$S_Z$ value in each holonomy does not evolve dynamically, therefore
remains non-zero if chosen to be non-zero initially.  It remains to
be seen whether this happy state of affairs will persist to a new
formalism with real eigenvalues for the area operator.

\appendix
\section{Details of the BPR calculation}
\renewcommand{\theequation}{\Alph{section}\arabic{equation}}
\setcounter{equation}{0}

     This appendix solves \eqs{3.5a}{3.5b} for the connection and
tetrads obeying BPR
symmetry, the invariance group for unidirectional plane
gravitational waves.  When setting up a complex connection
formalism, it is necessary to  choose three  phases: when
defining the Lagrangian at the four-dimensional level, one must
choose  the duality phase
$\delta$ and the phase of $\epsilon_{TXYZ}$ (see for example
\eq{3.6}); and an additional
phase comes in when rewriting the four-dimensional formalism  in
3+1 canonical
form..  These phases are explained in appendix A of II, and I use
the same phase choices here as in that paper.  I Begin at the
four-dimensional level by  solving
\eq{3.5a} for the Lorentz transformation L.
\be
     L_{I'I} = - \partial_{\beta} \xi^{\alpha}e_{\alpha I'}
e^{\beta}_I.
\label{eq:a.1}
\ee
I have dropped a $\xi^{\lambda}\partial_{\lambda}$\  term which
is
zero because $\xi^{(c)}$,
\eq{3.2}, is a linear combination of $\partial_x,\  \partial_y,\
and\ \partial_v$, all of which
annihilate $e^{\alpha}_I$.  In Rosen gauge, from \eq{3.2},
\be
     \partial_{\beta} \xi^{\alpha} = \delta^u_{\beta}g^{c\alpha}
-
                              \delta^c_{\beta}g^{u\alpha}.
\label{eq:a.2}
\ee
Therefore
\be
     L_{I'I} = -e^c_{I'}e^u_I + e^u_{I'}e^c_I.
\label{eq:a.3}
\ee
L is antisymmetric, as it should be.   \Eq{3.5a} determines L,
\eq{a.3}, but otherwise imposes
no new constraints on the
tetrads beyond those already imposed by $\partial_x,\
\partial_y,\
and\ \partial_v = 0$.

     Next consider \eq{3.5b}.  In principle, one should be able to
determine all the constraints on
the \A{A}{a} by solving these equations directly, but they are
awkward, and it is easier to adopt
an indirect approach.   Given the tetrads, compute the Lorentz
connection
$\omega^{IJ}_{\alpha}$; then compute
$^{(4)}A$\ , which is just the self-dual version of $\omega$.  In
this way one finds that many
components of $^{(4)}A$\   are identically zero.  When this
information is inserted into
\eq{3.5b}, that
equation reduces to the  trivial statement 0 = 0, for most values
of the indices; for a small
number of index values the equation is non-trivial and can be
solved with moderate effort.

     The equation relating $\omega$ to the tetrads is
\bea
     \omega_{ija} &\equiv & e_{iI}e_{jJ}\omega^{IJ}_a \nn
               &=&  [-g_{aj,i} + g_{ai,j} +
            e_{jK}\stackrel{\leftrightarrow}{\partial_a}e_i^K]/2.
\label{eq:a.4}
\eea
From this equation, at least one of i, j, or a must be u, since
derivatives with respect to x,y,v are zero.
Further, the tetrad matrix in Rosen gauge is 2x2 block
diagonal, with the zt to ZT block
containing constants only.   This implies that {\it at most} one
of i, j, or a must be u.   After
stripping off the (block diagonal) tetrads, one finds that the only
non-zero $\omega$ are
\be
     \omega^{XY}_u,  \omega^{VA}_a,
\label{eq:a.5}
\ee
where a = x,y only and A = X,Y only.   Now compute  $ ^{(4)}A$ from

$\omega$, using
\be
        2 ^{(4)}A^{IJ}_a = \omega ^{IJ}_a  + i\delta
(\epsilon^{IJ}_{..MN}/2
                         \epsilon_{TXYZ}) \omega ^{MN}_a,
\label{eq:a.5'}
\ee
where the duality eigenvalue $\delta \epsilon_{TXYZ}$ equals +1,
given my conventions $\delta = \epsilon_{TXYZ} = =1$.   The only
non-zero $^{(4)}A$\  components are
\be
     ^{(4)}A^{XY}_u,  ^{(4)}A^{TZ}_u,  ^{(4)}A ^{V+}_a.
\label{eq:a.6}
\ee
At first glance one might think this list is too short; there
should be  non-zero $^{(4)}A
^{U\pm}_a$ as well, because the $\epsilon^{IJ}_{..MN}$ term in
\eq{a.5'} will map the VA
indices on $\omega^{VA}_a$ into UB (B = X,Y or $\pm$).
Duality maps VX into VY, {\it not} UY, however, because of the
identities
\be
     \epsilon_{TXYZ} = \epsilon_{UVXY} = \epsilon^{VX}_{..VY},
\label{eq:a.7}
\ee
etc.\ , where the last, mixed index tensor with two V's is the
one
which occurs in duality relations.  This explains why there are no
$^{(4)}A ^{U\pm}_a$ ; to see
what happened to $^{(4)}A ^{V-}_a$ , note
\bea
        2 ^{(4)}A ^{V\pm}_a &=&  \omega ^{V\pm}_a  + i\delta
(\epsilon^{V\pm}_{..MN}/2
                                    \epsilon_{TXYZ}) \omega ^{MN}_a
\nn
               &=& \omega ^{V\pm}_a  + i\delta
(\epsilon^{V\pm}_{..V\mp}/
                                   \epsilon_{TXYZ}) \omega
^{V\pm}_a \nn
               &=& \omega ^{V\pm}_a (1 \pm \delta).
\label{eq:a.7'}
\eea
For my phase choice $\delta = +1$, $^{(4)}A ^{V-}_a$
vanishes.

     From \eq{a.6}, there is no need to consider $\alpha = v$ in
\eq{3.5b}.  I consider first $\alpha =
a = x,y$.  The first term in \eq{3.5b} is a linear combination of
$\partial_x,\
\partial_y,\ and\ \partial_v$, which vanishes for any $\alpha$.
The second term vanishes from
\eq{a.2} and the absence of any $\lambda = v$ component of $
^{(4)}A$.  Then \eq{3.5b} collapses to
\be
     0 = 0 + 0 + {\cal L}^I_{.I'}\ ^{(4)}A^{I'J}_a +
               {\cal L}^J_{.J'}\ ^{(4)}A^{IJ'}_a - 0.
\label{eq:a.8}
\ee
From \eqs{a.3}{3.6}, the only non-zero elements of {\cal L} are
\be
     {\cal L}_{UX} = i{\cal L}_{UY} = (e^c_X + ie^c_Y)/2 =
e^c_+/\sqrt{2};
\label{eq:a.9}
\ee
or
\be
      {\cal L}_{U+} = e^c_+; {\cal L}_{U-} = 0.
\label{eq:a.9a}
\ee
Inserting this and \eq{a.6} into \eq{a.8}, one finds that all
the $\alpha = a = x,y$ equations are trivially 0 = 0.

     Finally,  consider $\alpha = u$.  The first term in \eq{3.4}
vanishes as before.  The only IJ
index pair which does not give 0 = 0 is IJ = VA, A = X,Y only,
which gives
\be
     0 = 0 + \partial_u\xi^{\lambda}\ ^{(4)}A^{VA}_{\lambda} +
           {\cal L}^V_{.I'}\ ^{(4)}A^{I'A}_u +
                {\cal L}^A_{.J'}\ ^{(4)}A^{VJ'}_u -
                \partial_u{\cal L}^{VA}.
\label{eq:a.12}
\ee
Use \eq{a.2} to simplify the first term; use \eq{a.9a} to
simplify
the remaining terms and to
show that the A = - equation is trivial.  Then
\bea
     0 &=& 0 + g^{ca}\A{V+}{a} - 2e^c_+ \A{-+}{u} + \partial_u
e^c_+ \nn
     &=& g^{ca}\A{V+}{a} - 2e^c_B \A{B+}{u} + \partial_u e^c_+.
\label{eq:a.13}
\eea
I have used the duality relation $A^{VU} = -iA^{XY}$, and $
A^{-+}
= iA^{XY}$.  On the
second line, recall that a ``plus'   index always pairs with a
``minus'   index
to form the two-dimensional dot
product: $e^c_B A^{B+} = e^c_+ A^{-+} + 0$.   The $A_u$ field in
\eq{a.13} is a linear combination of $A_z$ and $A_t$ fields, and
the $A_t$ fields are non-dynamical Lagrange multipliers for the
Gauss constraints.  I therefore eliminate the $A_u$ field in order
to obtain a constraint on the dynamical field $\A{V+}{a}$.
From duality and \eqs{a.4}{a.5} for $\omega$,
\bea
     2e_{aB} ^{(4)}A^{B+}_u& =&e_{aB}[ \omega^{B+}_u +
i(\delta/2 \epsilon_{TXYZ})
                               \epsilon^{B+}_{MN}\omega^{MN}_u ]
\nn
               &=& [e^{j+}\omega_{aju} + 0] \nn
               &=&e^{j+}
[e_{jK}\stackrel{\leftrightarrow}{\partial_u}e_a^K ]/2 .
\label{eq:a.14}
\eea
I solve \eq{a.13} for $A^{VA}$ and insert \eq{a.14}.
\bea
     ^{(4)}\A{V+}{a} &=&
e^{j+}[e_{jK}\stackrel{\leftrightarrow}{\partial_u}e_a^K]/2
               - g_{ca}\partial_u e^c_+ \nn
          &=&e^{j+} [2e_{jK}\partial_u e_a^ K - \partial_u g_{ja}
]/2 - \partial_u e_{a+} +
                    \partial_u g_{ca} e^c_+ \nn
          &=&  \partial_u g_{ca} e^c_+ /2.
\label{eq:a.15}
\eea
The right-hand side of \eq{a.15} is proportional to the part of
\A{V+}{a} which contains no
$i\delta$ factor.
\bea
     2\mbox{``Re''}  ^{(4)}\A{V+}{a}&\equiv& (\omega^{VX}_a +i
               \omega^{VY}_a )/\sqrt{2} \nn
          &=& e^{iV}(e^{Xj}+ie^{Yj})\omega_{ija}/\sqrt{2} \nn
          &=&e^{+j}\partial_u g_{aj}/2.
\label{eq:a.16}
\eea
Therefore
\bea
      0 &=&  - ^{(4)}\A{V+}{a} + 2\mbox{``Re''  }  ^{(4)}\A{V+}{a}
\nn
     &=&  ^{(4)}\A{V+}{a}, \mbox{\ for\ } \delta = -1.
\label{eq:a.17}
\eea
The second line means that the first line is the four dimensional
connection computed with the
opposite choice for the duality eigenvalue, $\delta = -1$ rather
than $\delta = +1$.

     This result is very easy to transform from Rosen to a
general
gauge in the z,t sector, since the ``minus'' and ``a'' indices in
the x,y sector
remain invariant under such a transformation.  In order to
maintain the gauge conditions \eqs{3.3}{3.4} on the tetrads, it is
mecessary to combine any four-dimensional diffeomorphism (z,t)
$\rta$ (z't')  with
a Lorentz transformation, as at \eq{3.5a}.  A short calculation
shows that a very simple Lorentz
transformation $L^T_{.Z} = \partial t'/ \partial z$ will maintain
all the gauge conditions of
\eqs{3.3}{3.4}.  For transforming $ ^{(4)}A$ one needs the
corresponding ${\cal L}$; the only
non-zero matrix elements will be ${\cal L}^T_{.Z} $ and ${\cal
L}^X_{.Y}$; or equivalently ${\cal
L}^U_{.U}$, ${\cal L}V^V_{.V}$, and ${\cal L}^{\pm}_{.\mp}$.   Thus
the
coordinate transformation to the general gauge amounts to a Lorentz
transformation which multiplies \eq{a.17} by an overall factor.
\be
     ^{(4)}\!A^{V+'}_{a} = {\cal L}^V_{.V} {\cal L}^+_{.-}\!
               ^{(4)}\!A^{V+}_{a}, \mbox{\ for\ } \delta = -1,
\label{eq:a.11}
\ee
where every ${\cal L}$ and every $ ^{(4)}A$ is to be calculated
using the $\delta$ = -1
convention.  From \eq{a.11}, the quantity in \eq{a.17} vanishes in
every gauge.
Similarly, from \eq{a.6}, the following quantities vanish in every
gauge:
\bea
     0& =&  - ^{(4)}\A{U+}{a} + 2\mbox{``Re''\ }  ^{(4)}\A{U+}{a}
\nn
     &=& ^{(4)}\A{V-}{a} \nn
     &=& ^{(4)}\A{U-}{a} .
\label{eq:a.18'}
\eea
The  \eqs{a.18'}{a.17} may be rewritten as
\bea
     0 &=& ^{(4)}\A{T+}{a} - 2{\mbox``Re' }\  ^{(4)}\A{T+}{a}\nn
     &=& ^{(4)}\A{Z+}{a} - 2{\mbox``Re' }\  ^{(4)}\A{Z+}{a}\nn
     &=& ^{(4)}\A{T-}{a} \nn
     &=& ^{(4)}\A{Z-}{a},
\label{eq:a.18}
\eea
where every connection in \eq{a.18} is evaluated using the $\delta
= +1$ convention.  For the opposite duality
convention, $\delta = -1$, exchange $ (+
\leftrightarrow -) $ everywhere in
\eq{a.18}.  For $\delta = +1$ but left-moving rather than right
moving
waves, again exchange $ (+ \rla -) $ in \eq{a.18}.

     So far the calculation has been carried out entirely at the
four-dimensional level.  The four-dimensional connection
$^{(4)}A $ is related to the usual 3+1 connection A by the
following equation from Appendix A of II.
\be
     \A{S}{a} = - \epsilon_{MNS} ^{(4)} \A{MN}{a}.
\label{eq:a.19}
\ee
Then the  four-dimensional \eq{a.18} implies the 3+1
dimensional equations
\bea
     0 & = & \A{-}{a} ; \nn
     0 & = & -\A{+}{a} + 2 \mbox{``Re''} \A{+}{a}.
\label{eq:a.20}
\eea
Again, exchange $ (+ \rla -) $ for the opposite duality
convention or left-moving waves.

\section{Kinematics of the \A{\pm}{a} fields: spin}
\setcounter{equation}{0}

     From paper IV , the integral
\be
     L_Z =i \int dz [\E{y}{I} (\A{I}{x} - Re \A{I}{x})
                    - (x \leftrightarrow y)]
\label{eq:c9}
\ee
gives the total spin angular momentum of the wave, and is a
constant of the motion \cite{IV}.  The integral is over the entire
wave
packet, that is from $z_l $ to $z_r $.  As in II, the fields and
Weyl tensor components which produce transverse displacements of
test particles are
assumed to vanish at the boundaries, with
support only in the region $z_l < z < z_r $.  It is at first sight
surprising that any conserved quantity associated with the Lorentz
group should be given by a volume integral (integral over z) rather
than by a surface term (term evaluated at the endpoints $z_l $ and
$z_r $).  However, in the one-dimensional planar
case, the extensive
gauge-fixing in the x,y plane removes all gauge freedom, except
for  rigid rotations around z, and the x,y sector of the theory
resembles special relativity rather than general relativity.

     This appendix rewrites the integrand of $L_Z$ in  terms of the
unidirectional fields defined at \eq{2.9a}, in order to understand
the spin content of the latter.  Introduce triads $ e^A_a $ and
inverse triads $ e^a_A $, and write the integrand of $ L_Z $ as
\bea
     \E{y}{I} Im \A{J}{x} - (x \rla y) & = & [(e^y_J e_{Kx}
                         - (x \rla y)] \E{a}{K} Im \A{J}{a} \nn
               & = & [(e^y_{[J} e_{K]x} + e^y_{(J} e_{K)x}
                         - (x \rla y)] \E{a}{K} Im \A{J}{a}.
\label{eq:c14}
\eea
On the last line the term antisymmetric in J,K is proportional to
\E{a}{K} Im \A{J}{a} $ \epsilon_{JK} $.  This expression is part
of the Gauss constraint $\partial_z \E{z}{Z} + \E{a}{K} \A{J}{a}
\epsilon_{JK} = 0 $, which implies $ \E{a}{K} Im \A{J}{a}
\epsilon_{JK} = 0 $.  Hence the term antisymmetric in J,K can be
dropped.  The term symmetric in J,K can be expanded in O(2)
eigenstates, keeping in mind that every + index must be
contracted with a - index.  The $J \neq K $ terms are
proportional to
\bea
     e^y_+ e_{-x} + e^y_- e_{+x}  & = & e^y_B e_{Bx}
                                        \nn
                                   & = & \delta ^y_x \nn
                                   & = &  0.
\label{eq:c15}
\eea
Hence these terms can be dropped also.  The surviving terms are
products of tensors with J = K, therefore helicity $ \pm 2 $ in
the local Lorentz frame, a reassuring result.
\bea
     L_Z& =& -\int dz [ e^y_+ e_{+x} \E{a}{-} \mbox{ Im }\A{-}{a}
               + e^y_- e_{-x} \E{a}{+} \mbox{ Im }\A{+}{a} - (x
                                   \rla y) ] \nn
     &=& i \int dz \{ e^y_+ e_{+x} \E{a}{-}[\A{-}{a} + (\A{-}{a}
          - 2 \mbox{ ``Re'' } \A{-}{a}) \nn
     & &   + e^y_- e_{-x} \E{a}{+} [\A{+}{a} + (\A{+}{a}
          - 2 \mbox{ ``Re'' } \A{+}{a}) \} - (x  \rla y).
\label{eq:c16}
\eea
On the last line I have written $L_Z$ in terms of the
unidirectional BPR fields introduced at
\eq{2.9a}  This expression for
gravitational spin angular momentum,  possesses the same
coordinate times momentum
structure as the corresponding expression for electromagnetic
spin angular momentum:
\bea
     \vec{L}_{em} & = & (1/4 \pi)\int d^3x [\vec{E} \times
                              \vec{A}   \nn
          & = & -\int d^3x [\vec{\Pi} \times \vec{A} ]
\label{eq:c17}
\eea
That is , one can interpret the unidirectional quantities \Etld A
and \Etld (A - 2 ``Re'' A) in
\eq{c16} as momenta associated with waves of definite helicity.
This parallel with QED does
not extend too far: these ``momenta'   have nothing like
free-field commutation relations with
each other, or with the triad ``coordinates''  .

     It is now clear why one wants the two combinations
\E{a}{+}\A{+}{a}  and (-\A{-}{a}
+ 2 ``Re''  \A{-}{a}) \E{a}{-} to vanish: these two constraints
remove left-moving helicity $\pm
$ 2 contributions from $ L_Z $.  Why must the remaining, helicity
zero combinations vanish?
The helicity zero combinations are  \E{a}{-}\A{+}{a} and
(-\A{-}{a} + 2
``Re''  \A{-}{a}) \E{a}{+}.
They are complex conjugates of each other, so that by adding and
subtracting them from each
other one gets pure imaginary and pure real constraints
\bea
     0 & = &\E{a}{B} (\A{B}{a} - Re\A{B}{a}) -
               i\epsilon_{AB}Re\A{A}{a}\E{a}{B};
\label{eq:c18i}                          \\
     0 & = &- i\epsilon_{AB} (\A{A}{a} - Re\A{A}{a}) \E{a}{B}
                          + \E{a}{B} Re \A{B}{a}.
\label{eq:c18r}
\eea
Now consider the classical equation of motion
\bea
     0 & = & -i \E{z}{Z}_t - \delta H / \delta \A{Z}{z} \nn
     & =&  -i \E{z}{Z}_t  - \E{a}{B} \A{B}{a}.
\label{eq:c19}
\eea
On the second line I use the Hamiltonian of \eq{4.5}.  I
also use the unidirectionality assumption (for the first time in
this section; $L_Z$ is the spin operator also for the scattering
case)  and evaluate the Hamiltonian in Rosen (or at
least conformally flat) gauge.  The metric in a general gauge has
the form
\be
     ds^2 = [(-(N')^2 +(N^z)^2)dt^2 + 2N^z dzdt +
dz^2]g_{zz} + \mbox{x,y sector},
\label{eq:c19'}
\ee
so that to obtain conformal gauge, one must take $N^z$ = 0
and N' = 1 in the Hamiltonian, where $N^z$ is the shift and N' is
the renormalized lapse defined following \eq{4.5}.   From the real
part of \eq{c19}, the \Etld\ Re A term in
\eq{c18r} vanishes.  The rest of
this equation is just the imaginary part of the Gauss constraint,
$\partial_z \E{z}{Z} +
\epsilon_{AB} \A{A}{a} \E{a}{B} = 0 $, and vanishes also.  This
leaves \eq{c18i}.  The $
\epsilon $ Re A \Etld term may be simplified using the real part
of the Gauss constraint; the \Etld
(A - Re A) term may be simplified using the imaginary part of the
equation of motion, \eq{c19}.
The result is simply
\be
     0 = -i (\partial_t + \partial_z ) \E{z}{Z} ,
\label{eq:c20}
\ee
in any conformally flat gauge.
This equation is discussed further at \eq{3.9} of section II.

\section{The ADM energy}
\setcounter{equation}{0}

     It is a worthwhile exercise to express the ADM energy in
terms of BPR operators.  In the
usual three-space dimensional case, the Hamiltonian expressed in
terms of the original ADM
variables is the sum of a volume integral plus a surface term
$H_{st}$
\cite{deW, RTei},
\be
     H = \int d^3x [NC_{sc} + N^iC_i ]  + H_{st}.
\label{eq:d.1}
\ee
In the classical theory, the ADM energy is just the surface term,
since the constraints in
the volume term must vanish everywhere when the solution obeys the
classical equations of motion.   Often one says that in the quantum
case the ADM energy is just the surface term also, but this is not
quite right, as we shall see in a minute.

     In the planar, one-space dimensional case, the
expression for the Hamiltonian
in terms of Ashtekar variables looks superficially much the same as
\eq{d.1} \cite{II},
\be
     \int dz [N'C_{sc} + N^zC_z + N_GC_G]  + H_{st},
\label{eq:d.2}
\ee
except for the additional Gauss constraint, and the prime on N'.
(The prime means I have
renormalized the usual Ashtekar lapse by absorbing a factor of
\E{z}{Z} into the lapse, as
explained in II.)   In both one and three spatial dimensions, one
might be tempted to drop the volume terms, in the quantum
mechanical case, because the constraints $C_i$ are required to
annihilate the wavefunctional.  However, the statement that
the scalar constraint (say)
annihilates the wavefunctional means, not $C_{sc}\psi = 0$, or even
$\int dz C_{sc}\psi = 0$,
but rather
\be
     \int dz \delta N' C_{sc}\psi = 0,
\label{eq:d.3}
\ee
where $\delta N' $ is a small change in the lapse.
The arbitrary  change  $\delta N' $ must preserve the boundary
condition   at spatial infinity, $
N' \rta 1$ .  Hence  $\delta N' \rta 0$ there.  On the
other hand, when $C_{sc} $ occurs in the Hamiltonian of \eq{d.2},
it is multiplied by N' rather than  $\delta N' $.  There is no need
for N' to vanish at the boundaries (in fact it becomes unity
there).  Now suppose the evaluation of the action of $C_{sc}$ on
$\psi$ reqires an integration by parts with respect to z.  In
\eq{d.3}, when the constraint acts upon $\psi$, surface terms at $z
= z_b$ will vanish because of the boundary condition on $\delta
N'$.  When H acts upon $\psi$, however, the $C_{sc}$ in H is
smeared by N' rather than $\delta N'$; the former does not vanish
at boundaries.  Consequently the volume term can contribute to the
ADM energy in the quantum case.  In the planar case both the Gauss
and scalar constraints in the volume term can contribute to the ADM
energy; neither N' nor $N_G$ is required to vanish at the
boundaries.  In the usual 3+1 dimensional case with flat space
boundary conditions at infinity, only the scalar constraint in the
volume term can contribute; the remaining constraints are smeared
by $N_i$ which are required to vanish at spatial infinity.

     For the plane wave case, the  surface terms in the Hamiltonian
were computed  in section 4 of II.
\bea
     H_{st}
&=&-\epsilon_{MN}\E{b}{M}\A{N}{b}\mid_{z_l}^{z_r}
                         \nonumber \\
          &=&I[\E{b}{-}\A{+}{b} - \E{b}{+}\A{-}{b}]
                         \nonumber \\
          &=& i\hbar[\E{b}{-}\delta /\delta \E{b}{-}
                     - \E{b}{+}\delta /\delta
\E{b}{+}]_{z_l}^{z_r}
\label{eq:d.4}
\eea
To simplify the boundary term quoted in II, I have invoked the
boundary conditions $N^z \rta 0$,
$N' \rta 1$ on the shift and renormalized lapse.   Evidently the
ADM energy contains the BPR
operators which are sensitive to the long-range scalar potential,
which suggests that these
operators may play a role even in the presence of waves which are
not unidirectional.

     The operator of \eq{d.1} gives a finite result when applied to
the solutions of II-III;
there is no  need
to renormalize.    However, the solutions are not eigenfunctions of
this operator.   A typical
solution involves n integrations $dz_i$ over the locations of the
n $\E{a}{A}(z_i)$ operators
contained in the wavefunctional, and the ADM operator acts as a
``lowering operator'' ,
removing one integration.   Hence an eigenfunction would have to be
an infinite sum over
wavefunctionals of all possible values of n.  It is beyond the
scope of this paper to investigate the
finiteness of the norm of such a sum.

\section{The transverse Weyl criterion}
\setcounter{equation}{0}
1.  Classification of Weyl tensors.
\medskip

     The Weyl tensor is the part of the Reimann tensor which can
be
non-zero even in empty
space, and certain of its components induce transverse vibrations
when inserted into the equation
of geodesic deviation \cite{SzWeyl}.  It is therefore a natural
object to work
with when constructing a criterion
for the presence of radiation \cite{Zak}.  The construction
proceeds in two
steps.  The first step is a straightforward
mathematical problem: classify Weyl tensors using their algebraic
properties.  In the second step, one uses physical arguments to
determine the
Weyl class(es) most closely associated with radiation.

     To begin with the mathematical problem, there are 10
independent real
components of the Weyl tensor, and
from these one can construct 5 independent complex components
which
have simple duality
properties.
\bea
     {\cal C}_{abcd}& =& [C_{abcd} + i(\delta/2\epsilon_{TXYZ})
\epsilon_{abmn}
C^{mn}_{cd}]/2 ; \label{eq:e1} \\
     {\cal C}_{abcd}& =& i(\delta/2\epsilon_{TXYZ})
\epsilon_{abmn}{\cal
C}^{mn}_{cd}.
\label{eq:e2}
\eea
Lower case Roman indices a,b,c,$\ldots$ are global; upper case
Roman indices A,B,C,$\ldots$
are local Lorentz. $ \epsilon_{abmn}$ is the totally
antisymmetric
global tensor, while
$\epsilon_{TXYZ}$ is the corresponding local Lorentz quantity,
the
Levi-Civita constant tensor.
The duality eigenvalue is $\delta/\epsilon_{TXYZ} = \pm 1$.
There
is another Levi-Civita
tensor hidden in the $ \epsilon_{abmn}$,
\[
      \epsilon_{abmn} = e_a^A e_b^B e_c^C e_d^D
\epsilon_{ABCD};
\]
therefore $\epsilon_{TXYZ}$ and its associated sign convention
drop
out after the conversion to
Ashtekar variables and the 3 + 1 splitup.  The final 3+1
Hamiltonian contains only the phase
$\delta$.  My convention is  $\delta = +1$, but in this paper all
results are stated in a manner which facilitates a
conversion to the opposite convention.   Of course the
combinations  with simple duality
properties also have simple transformation properties in the
local
Lorentz frame, which is why
one chooses to work with ${\cal C}$ rather than C, when
attempting
a classification.

     Petrov was the first to classify Weyl tensors by their
algebraic properties \cite{Petrov}; but for
present purposes the equivalent classification scheme due to
Debever  \cite{Debvec,peeling}  is more convenient.  A
null vector k is said to be a principal null vector (Debever
vector) of   ${\cal C}$ if
\be
     k_{[a}{\cal C}_{b]mn[c}k_{d]}k^mk^n = 0; k_ak^a = 0.
\label{eq:e3}
\ee
Debever proved that a Weyl tensor can have up to four distinct
principal null vectors, and he
classified Weyl tensors by the number of degeneracies among these
vectors.   If [1111] denotes the Weyl tensors which have 4
distinct
Debever vectors, [211] the
Weyl tensors with two vectors degenerate and the rest distinct,
etc.\ , then the five classes are
[1111], [211], [22], [31], and [4].  (The corresponding five
Petrov
classes are I, II, D, III, and N,
respectively.)

     Now suppose k is a Debever vector obtained by solving
\eq{e3}.  Make it one leg of a
null tetrad k, l, m, \={m}.   Choose the Z axis of a local free
fall frame so that k and l have spatial components
along $\pm Z$, while m and \={m} are transverse.
\bea
     -k_a l^a &= & m_a \bar{m}^a = 1; \nn
     k_a k^a &=& l_a l^a = m_a m^a \nn
          &=& k_a m^a = l_a m^a = 0.
\label{eq:e5}
\eea
{\cal C} may be expanded in this basis, and (not surprisingly)
one
gets five possible terms.
\bea
     {\cal C}_{abcd} &=& C_1V_{ab}V_{cd} \nn
               & & + C_2(V_{ab}M_{cd} + M_{ab}V_{cd}) \nn
               & & + C_3(M_{ab}M_{cd} - U_{ab}V_{cd} -
V_{ab}U_{cd})\nn
               & & +C_4(U_{ab}M_{cd} + M_{ab}U_{cd}) \nn
               & & + C_5U_{ab}U_{cd},
\label{eq:e6}
\eea
where
\bea
     V_{ab} &=& 2k_{[a}m_{b]}; \nn
     U_{ab} &=& 2l_{[a}\bar{m}_{b]}; \nn
     M_{ab}&=& 2k_{[a}l_{b]} + 2m_{[a}\bar{m}_{b]}.
\label{eq:e7}
\eea
The five combinations in \eq{e6} are the only ones allowed by the
duality convention $\delta =
+1$.  The expansion for the opposite duality convention may
be obtained from \eqs{e6}{e7}
by interchanging m and \={m} in \eq{e7}.  \Eq{e6} is essentially
the expansion given by Szekeres \cite{SzWeyl},
after a relabeling of the basis vectors $(k,l,m,\bar{m}) \rta (k,-
m,t,\bar{t})$.   Since the expansion
treats k and l quite
symmetrically, it is valid also for the case that l, rather
than k is the principal null vector.

     At this point one turns from the mathematical to the
physical:
which Petrov/Debever class(es), or which term(s) in \eq{e6}, are
most closely associated with radiation?  Consider first which of
the five tensors in \eq{e6}  distorts a cloud of test
particles in the manner expected for gravitational radiation.
Szekeres finds that only the $C_1$ and $C_5$ terms produce the
transverse displacements in the XY plane
characteristic of gravitational radiation in the linearized
theory.
$C_2$ and $C_4$ produce
longitudinal displacements in the XZ or YZ planes. $ C_3$
produces
a Coulomb (or tidal force) displacement:   $ C_3$  distorts a
sphere
of particles into an ellipsoid of revolution
with axis along Z.  These facts suggest that the $C_1$
and $C_5$ terms signal the presence of radiation.

     There is another set of arguments which suggest that the
$C_1$
and $C_5$ terms are closely associated with the presence of
radiation.  If the Weyl tensor contains {\it only} a $C_1$ or
$C_5$
term, the tensor is type N.  (A type
N tensor with k as principal
null vector contains only a $C_1$ term; a type N tensor with l as
principal null vector contains only a $C_5$ term .)  Type N is
closely associated with radiation.  Along characteristic curves,
when the metric is discontinuous, the discontinuity in
the Weyl tensor is type N \cite{Pirlect}.   In the linearized
theory, the tensor
associated with unidirectional
gravitational radiation is type N.  At large distances
from bounded sources, the surviving components of the Weyl tensor
are type N ("peeling theorem" \cite{peeling}).

     Although the $C_1$ and $C_5$ terms are closely associated
with
type N, it would be better to call these terms transverse Weyl
components, rather than type N components, since a
tensor which is not type N can nevertheless
contain $C_1$ or $C_5$ terms.   A [22] (type II) field contains
$C_1$ plus some admixture of
longitudinal component, while [1111] (type I) contains $C_1$,
$C_5$, and $C_3$  terms.  In a
theory as non-linear as general relativity, one can expect that a
collision between $C_1$ and
$C_5$ transverse waves will produce some $C_3$ (Coulomb)
component,
and the tensor will be
type I rather than type N.  In  asymptotic regions, after the
transverse wave has "outrun" its
Coulomb companion, presumably the tensor will revert to type N;
but
in general one should be
looking for $C_1$ and $C_5$, rather than type N or any other
specific Petrov class.  One should describe this radiation
criterion as the transverse Weyl criterion, rather than the type
N
criterion.

     I now construct operators which
project out the  $C_1$ and $C_5$ terms.
\bea
      C_1 &=& {\cal C}_{abcd}l^a\bar{m}^bl^c\bar{m}^d; \nn
      C_5 &=& {\cal C}_{abcd}k^am^bk^cm^d;             \nn
      C_3 &=& -{\cal C}_{abcd}l^a\bar{m}^b k^c m^d.
\label{eq:e8}
\eea
For completeness I have included also the expression for the pure
Coulomb component $C_3$.  The plane wave case  has no
longitudinal components; $C_2$ and $C_4 $ vanish identically.

\medskip
\noindent 2. Transition to Ashtekar variables.
\medskip

     At this point I specialize to the case of plane waves along
the Z axis.  By definition, the plane wave metric has two
hypersurface orthogonal null vectors which may serve as normals
to
right- and left-moving wavefronts U = (cT-Z)/$\sqrt{2}$ = const.
and
V = (cT+Z)/$\sqrt{2}$ =
const.  I
identify these normals with k (right-moving) and l (left-moving),
so that a small change in the wave phase will look like
\bea
     k_adx^a &=& (-dT + dZ )/\sqrt{2} = -dU; \nn
     l_adx^a &=& (-dT - dZ)/\sqrt{2} = -dV.
\label{eq:e9}
\eea
(Hypersurface orthogonality demands $k_adx^a \propto dU$, etc.;
the normalization conditions force the constants of
proportionality to be as shown in \eq{e9}; and the overall phase of
$k_a$ and $l_a$ is fixed by the requirement that $k^0$ and $l^0$ be
positive, i.e.\ future-pointing.)
Lower case x denotes a global coordinate; upper case (T, Z, U, V)
denotes a coordinate in a local Lorentz frame.  From the
expression
for the (inverse) tetrads, $e^A_b dx^b = dX^A$, k and l may be
identified with the tetrads
\bea
     k^a &=& -e^{aU} = +e^a_V; \nn
     l^a &=& e^a_U.
\label{eq:e10}
\eea
Similarly,
\bea
     m^a &=& e^a_+; \nn
     \bar{m}^a &=& e^a_-.
\label{eq:e11}
\eea
Evidently the quantities $C_i$ are (global scalars and) tensors
in a Local Lorentz frame.

     The \eq{e9} places quite a strong restriction on the null
basis, going beyond what is required to maintain the normalization
\eq{e5}.  The choice \eq{e10} certainly is not unique.  For
example, $k_a$ remains null if it is rescaled by an arbitrary
function.  (Simultaneously $l_a$ must be rescaled by the inverse
function, in order to maintain the normalization condition $-
k_al^a$ = 1).  Similarly m and \={m} may be rescaled.  The choice
\eqs{e10}{e11} facilitates calculations and  leads to highly
symmetric formulas for $C_1$ and $C_5$.  (In this basis, $C_1$ is
just $C_5$ with some plus and minus indices interchanged; see
\eq{e19} below).  For further discussion of the effect of choice of
basis, see the remarks following \eq{e19}.

     Conversion to the Ashtekar language is straightforward.
$\cal C$ is essentially the
(four dimensional) Ashtekar field strength, since the $\cal
C$ tensor is self-dual, and in empty space the Weyl tensor
is the full Riemann tensor.
\be
     ^{(4)}F^{AB}_{cd} = e^{Aa} e^{Bb} {\cal C}_{abcd} .
\label{eq:e12}
\ee
The four-dimensional field strengths $^{(4)}F$  may be replaced by
3+1 quantities
F by using standard formulas.
\bea
     ^{(4)}F^{TM}_{cd} &=& -i\sigma \delta F^M_{cd} /2; \nn
     ^{(4)}F^{MN}_{cd} &=& \sigma \epsilon_{MNS} F^S_{cd} /2.
\label{eq:e13}
\eea
M,N,S = space only.  $\sigma$ is a new phase which appears at the
3+1
reduction step.  I choose $\sigma = -1$, for reasons explained in
Appendix A of II.  This phase ( unlike $\delta$ ) merely changes
the overall sign of the $C_i$,
and I shall not keep track of the $\sigma $ dependence in the
future. Useful corollaries of
\eq{e13} are
\bea
     ^{(4)}F^{V\pm}_{cd} &=& (i/\sqrt{2}) F^{\pm}_{cd}[(\delta
\pm
1)]/2;\nn
     ^{(4)}F^{U\pm}_{cd} &=& (i/\sqrt{2}) F^{\pm}_{cd}[(\delta
\mp
1)]/2.
\label{eq:e14}
\eea
 If tetrads \eqs{e10}{e11} and field strengths \eq{e14} are
inserted into \eq{e8} for $C_1$,
the result for $\delta = +1$ is
\be
     C_1 = i[-F^+_{cd}e^c_T + F^+_{cd}e^c_Z]e^d_+/2.
\label{eq:e15}
\ee
(For $\delta = -1$ replace + by -.)  For any metric, typically
the
Lorentz boosts are gauge-fixed
by demanding that three of the tetrads vanish:
\be
      e^t_M = 0, M = space.
\label {eq:e16a}
\ee
For the special case of the plane wave metric, the gauge-fixing
of
the XY Gauss constraint and
xy spatial diffeomorphism constraints imply that four more
tetrads
vanish.
\be
     e^z_X = e^z_Y = e^x_Z = e^y_Z = 0.
\label{eq:e16b}
\ee
The tetrad matrix reduces to two 2x2 subblocks which link x,y to
X,Y (or\ $ \pm$)  and z,t to Z,T.  Therefore the first term in
\eq{e15} ( and only the first term) contains an $ F^+_{td}$ term,
d = x or y, with  unacceptable time derivatives of the
"coordinate"
$A^+_d$.  I eliminate this term
using the classical equations of motion, which are
\be
      ^{(4)}F^{AB}_{cd}e^c_A = 0,
\label{eq:e16}
\ee
or after 3+1 splitup, and setting B = +,
\bea
     0 &=& ^{(4)}F^{A+}_{cd}e^c_A \nn
      &=& ^{(4)}F^{-+}_{cd}e^c_+ + ^{(4)}F^{U+}_{cd}e^c_U +
                ^{(4)}F^{V+}_{cd}e^c_V \nn
          &=& F^+_{cd}e^c_V \nn
     &=& F^+_{cd}e^c_T + F^+_{cd}e^c_Z.
\label{eq:e17}
\eea
On the second line the $ ^{(4)}F^{U+}$ term vanishes because of
\eq{e14}, and the $
^{(4)}F^{-+}_{cd}e^c_+$ may be dropped because at the next step
the
entire term will be
contracted with $e^d_+$.  When \eq{e17} is inserted into
\eq{e15}, the result is
\be
     C_1 = i F^+_{cd}e^c_Z e^d_+.
\label{eq:e18}
\ee
This is not quite Ashtekar form, because the triads must be
densitized.  Also, the c index can
equal z only, because of the gauge conditions \eqs{e16a}{e16b}.
The final result is (for $C_5$
and $C_3$ also, since they are calculated similarly)
\bea
     C_1 &=& i F^+_{zd}\E{d}{+}/\Etwo; \nn
     C_5 &=& i F^-_{zd}\E{d}{-}/\Etwo  \nn
     C_3&=& F^Z_{xy}/2\E{z}{Z}.
\label{eq:e19}
\eea
\Etwo\  is the determinant of the 2x2 XY subblock of
the tetrad matrix.  The results for $\delta = -1$ are the same,
except for overall
phases, and interchange of + and -
everywhere.

     In the case that the wave is unidirectional, the results
\eq{e19} are consistent with the BPR constraints.  For example,
if
the wave is right-moving, then the principal vector is k,
associated with the tensor $C_1$.  From \eq{3.7a},
\A{-}{a} vanishes, implying that ($C_1$ is finite, while) $C_5$
vanishes.

     The $C_i$ of \eq{e19} were calculated in a specific basis; in
the language of section III, they are not even kinematic, much less
physical.  In particular the factors of \Etwo\ in \eq{e19} are
basis dependent.  For example, suppose one shifts from the tetrad
basis, \eq{e10}, to a basis in which $k^a$ is affinely
parameterized.  (In the tetrad basis one has $k^b k_{a;b} = \lambda
k_a$, $\lambda \neq 0$.)  Then the \Etwo\ factor in $C_1$
disappears, replaced by a factor of \E{z}{Z}.  One could continue
to rescale $C_1$ in this manner, until it became density weight
unity; then it could be sandwiched between holonomies and
integrated over z to make it kinematic.  Presumably one would have
to tolerate some degree of non-polynomiality in the final result.

     Although it would not be hard to make the transverse operator
$C_1$ kinematical, it is unikely that $C_1$ could be  made
physical.  Gravitational radiation
is closely identified with transversality only in the the
linearized theory.  In the full classical theory, scattering of
two transverse waves produces a $C_3$ Coulomb component
\cite{Szmet}.  Presumably,
then, the commutator of any purely transverse operator with the
Hamiltonian will not be especially simple, even in the classical
theory.  This is one reason why the main body of the paper
concentrates on the BPR operators, rather than the Weyl tensor.
Although it is unlikely that any transverse criterion could be
made physical, a kinematical criterion for transversality should be
both feasible and useful.

\end{document}